\documentclass[12pt]{article}
\usepackage{amssymb}
\usepackage{amsfonts}
\usepackage{mathrsfs}
\usepackage{latexsym}
\usepackage{amsmath}
\usepackage{graphics}
\usepackage{graphicx}
\usepackage{sectsty}
\usepackage{setspace}
\usepackage{multirow}
\sectionfont{\normalsize\bf}
\subsectionfont{\rm\normalsize}
\def\be{\begin{equation}}
\def\ee{\end{equation}}

\def\half{{\scriptstyle {1\over 2}}}

\def\bn{{\boldsymbol n}}

\newcommand{\da}{\dot{a}}
\newcommand{\db}{\dot{b}}
\newcommand{\bpi}{\bar{\pi}}
\newcommand{\bs}{\bar{s}}

\numberwithin{equation}{section}
\begin{document}

\thispagestyle{empty}
\begin{flushright}
\end{flushright}
\baselineskip=16pt
\vspace{.5in}
{%\Large
\begin{center}
{\bf Conformal Supergravity Tree Amplitudes\\ from Open Twistor String 
Theory}
\end{center}}
\vskip 1.1cm
\begin{center}
{Louise Dolan and Jay N. Ihry}
\vskip5pt

\centerline{\em Department of Physics}
\centerline{\em University of North Carolina, Chapel Hill, NC 27599} 
\bigskip
%\centerline{\bf ldolan, jihry@physics.unc.edu}
\bigskip
\bigskip
\end{center}

\abstract{\noindent 
We display the vertex operators for all states in the conformal
supergravity sector of the twistor string, as outlined by
Berkovits and Witten. 
These include `dipole' states, which are pairs of supergravitons
that do not diagonalize the translation generators.
We use canonical quantization of the open string version of Berkovits,
and compute $N$-point tree level scattering amplitudes for gravitons,
gluons and scalars. We reproduce the Berkovits-Witten formula for maximal 
helicity violating (MHV) amplitudes (which they derived using path integrals),
and extend their results to the dipole pairs. We compare these 
trees with those of Einstein gravity field theory.
\bigskip

\vskip120pt email: ldolan, jihry@physics.unc.edu
\setlength{\parindent}{0pt}
\setlength{\parskip}{6pt}

\setstretch{1.25}

%\tableofcontents
\vfill\eject
\section{Introduction}

We pursue the tree amplitudes for graviton scattering 
in conformal gravity, described by twistor string theory.
The twistor string \cite{W} and its open string formulation \cite{B}
describe massless particles of ${\cal N}=4$ Yang-Mills theory coupled to 
conformal supergravity \cite{BW} in four-dimensional
Minkowski spacetime. 

Conformal gravity field theories \cite{BR, FT1}
provided early examples
of finite field theories of gravity \cite{FT2, RvN}.
They are not unitary theories,
but have interesting structure and continue to provoke comments about
possible uses \cite{Bender}.
Of course the conformal supergraviton states have zero norm, 
due to the lack of unitarity \cite{FZ}.  
Nevertheless, the equivalence of the twistor string with this
field theory system can be exploited to derive
conformal gravity tree level scattering amplitudes hard to  
access in the field theory.
We compute the gravity trees as a step toward learning how to 
decouple them in the twistor string. This would result in a perturbative
string theory for super Yang-Mills (with no 
tower of massive states), and the 
computational advantage one hopes for in a string theory vs. field theory
description. Various efforts towards a QCD string are discussed in
\cite{AZ}. 

We work in a spinor helicity basis \cite{PT}-\cite{CSW}, 
and compare the conformal
gravity tree amplitudes with those of Einstein gravity 
\cite{Berends}-\cite{BB}. The conformal gravity trees have fewer poles. 
We compute the conformal couplings in detail, as they should be important
in further study of the loop calculation \cite{DG1}.

Computation is done in the Berkovits open string version \cite{B}.  
We use the twistor string canonical quantization described 
in \cite{DG1,DG2}
and follow their notation. In section 2, we
give the vertex operators for all states in the conformal supergravity 
multiplets, as outlined by Berkovits and Witten \cite{BW}.
These include the dipole states, which form pairs of supergravitons,
where one state in each pair does not diagonalize the translation
generators, and is not a momentum eigenstate. 
We show all supergraviton states have zero norm in our basis.

In section 3, three-point scattering amplitudes 
for gluons and gravitons and scalars, with both one and two negative
helicities are calculated. 
We include cases for both members the dipoles,
and find a momentum derivative appearing in amplitudes for states that
do not diagonalize the translations generators. 
These amplitudes still have translational invariance.

In section 4, we extend our results to $N$-point tree level amplitudes
for these dipole pairs.
We reproduce the Berkovits-Witten formula 
for maximal helicity violating (MHV) amplitudes for the 
diagonal states, showing consistency of the canonical approach and the
path integral framework.
In section 5, we compute $N$-point conformal gravity MHV tree amplitudes
for a selection of gluons and supergravitons in the dipole 
pairs.  

% SECTION
\section{Vertex Operators and Canonical Quantization}

The world-sheet fields for twistor string theory are the
twistor fields $Y_I, Z^I$, $1\le I\le 8$, 
and the current algebra $J^A$ with central charge $28$.
In addition, there are ghost fields $b,c,u,v$, and world sheet gauge fields
all summarized in \cite{DG1}. 
The fields $Z^I$ have conformal spin zero and
are relabeled as four boson fields $\lambda^a,\mu^{\dot a}$,
$1\le a, \dot a \le 2$  and
four fermion fields $\psi^m$, $1\le m\le 4$. The conjugate variables
$Y_I$ have conformal spin one, as do the currents $J^A$.

The twistor field commutation relations follow from
\begin{align}
Z^I(\rho)Y_J(\zeta) = :Z^I(\rho)Y_J(\zeta): + \delta^I_J (\rho-\zeta)^{-1}.
\end{align}

\subsection{Vertex Operators}

The massless states of ${\cal N}=4$ conformal supergravity
consist of pairs of graviton supermultiplets (called dipoles), whose
vertex operators are $V_F(\rho)$, $V_{F'}(\rho)$ and
$V_G(\rho), V_{G'}(\rho)$;
in addition to spin $3/2$ supermultiplets, with vertex operators
$V_f(\rho)$ and $V_g(\rho)$.
Loosely following the notation of \cite{BW},
we list them in terms of homogeneous functions $f^I$, $g_I$,
of $Z^I$ in \break Table 1. For each vertex, $f^I$ and $g_I$ satisfy
\begin{align}{\partial\over\partial Z^I} f^I = 0, \quad
Z^I g_I = 0\end{align}
to ensure the vertex operators are primary with respect to 
the $U(1)$ current 
\begin{align}J(\rho) = -\sum_I :Y_I(\rho) Z^I(\rho):\end{align}
and the Virasoro current
\begin{align} L(\rho) = -\sum_I :Y_I(\rho) Z^I(\rho): - :u(\rho)v(\rho):
+ 2:\partial c(\rho) b(\rho): - :\partial b(\rho)c(\rho): + L^J(\rho).
\end{align}
Here $L^J(\rho)$ is the contribution from the current algebra.
The vertex operators have charge zero and conformal dimension one. 
The primed vertices correspond to states that do not diagonalize the
translation generators \cite{BW}.
%% TABLE
\begin{center}
\renewcommand{\tabcolsep}{1cm}
\renewcommand{\arraystretch}{1.5}
\vskip20pt
\begin{tabular}{| l | c |}
\hline
\hspace{1.25cm} {\bf Vertex Operator} & {\bf Helicities} \\ \hline
$V_F(\rho) = f^{\da}(Z(\rho))Y_{\da}(\rho)$ 
& $(2,\, \frac{3}{2},\, 1,\, \half ,\, 0)$  \\ \hline
$V_G(\rho) = g_a(Z(\rho))\partial \lambda^a(\rho)$ 
& $(0,\, -\half,\, -1,\, -\frac{3}{2},\, -2 )$ \\ \hline 
$V_{F'}(\rho) = f^a(Z(\rho))Y_a(\rho) + \hat{f}^{\da}(Z(\rho))Y_{\da}(\rho)
$ 
& $(2,\, \frac{3}{2},\, 1,\, \half,\, 0 )$\\ \hline
$V_{G'}(\rho)= g_{\da}(Z(\rho))\partial \mu^{\da}(\rho) + \hat{g}_a(Z(\rho))
\partial \lambda^a(\rho)$ & $(0,\, -\half,\, -1,\, -\frac{3}{2},\, -2 )$
\\ \hline
$V_f(\rho) = f^m(Z(\rho))Y_m(\rho) + \tilde {f}^{\da}(Z(\rho))Y_{\da}(\rho)$ 
& $( \frac{3}{2},\, 1,\, \half,\, 0,\, -\half )$\\ \hline
$V_g(\rho) = g_m(Z(\rho))\partial \psi^m(\rho) + \tilde{g}_a(Z(\rho))
\partial \lambda^a(\rho)$ & $(\half,\, 0,\, -\half,\, -1,\, -\frac{3}{2} )$ 
\\ \hline
$V_\Phi^A(\rho) = 
V_{\phi}(Z(\rho))J^A(\rho)$& $(\pm 1,\, 4(\pm \half), \, 6(0) )$  \\ \hline
\end{tabular}
\vskip10pt
\centerline{Table 1: Vertex operators and helicities for ${\cal N}=4$ conformal
supergravity and Yang-Mills theory}
\end{center}

We will define the homogeneous functions for each vertex operator, and 
discuss their properties. The states are labeled by helicities
and their representations under the $SU(4)$ $R-$symmetry (in bold).  
As a reminder, we first look at the ${\cal N}=4$ Yang-Mills
gluon vertex, 
\be
V_{\Phi}^A(\rho) = V_{\phi}(Z(\rho))J^A(\rho)
\ee
with
\begin{align}
V_{\phi}(Z(\rho)) &= \int\frac{dk}{k} \prod^2_{a=1}\delta(k\lambda^a(\rho) 
-\pi^a) e^{ik\bar\pi_{\dot b} \mu^{\dot b}(\rho)} \cr
& \quad \times \left[ A_{1} + k\psi^bA_b + \frac{k^2}{2} \psi^b\psi^c A_{bc} 
+\frac{k^3}{3!} \psi^b\psi^c\psi^dA_{bcd} + k^4 \psi^1\psi^2\psi^3\psi^4
A_{-1}\right]\cr
&= {1\over (\pi^1)^2}\delta\left({\lambda^2(\rho)\over \lambda^1(\rho)}
-{\pi^2\over\pi^1}
\right)\exp\left\{i{\mu^{\dot b}
(\rho)\bar\pi_{\dot b}\pi^1\over\lambda^1(\rho)}\right\}\cr
&\hskip5pt \times\big[A_+
+{\pi^1\over\lambda^1(\rho)}\psi^b(\rho)A_b+\left({\pi^1
\over\lambda^1(\rho)}\right)^2{1\over 2}\psi^b(\rho)\psi^c(\rho)A_{bc}
\cr&\hskip13pt +\left({\pi^1
\over\lambda^1(\rho)}\right)^3{1\over 3!}\psi^b(\rho)\psi^c(\rho)\psi^d(\rho)
A_{bcd} +\left({\pi^1
\over\lambda^1(\rho)}\right)^4\psi^1(\rho)\psi^2(\rho)\psi^3(\rho)\psi^4(\rho) 
A_-\big]\cr
\end{align}
where $\psi^b \equiv \psi^b(\rho)$ and $b,c,d$ are summed over.  
With use of the delta function $\delta (k\lambda^1(\rho) - \pi^1)$ to
perform the $k$-integration, this becomes
the vertex used by Berkovits and Witten, except they omit 
the $A_b$, $A_{bc}$, and $A_{bcd}$ terms \cite{DG1, B, BerkMotl, BW}.  
In that form, it is easy to see that the vertex operator
$V_\phi(Z^I(\rho))$ is homogeneous in $Z^I(\rho)$ of degree $p=0$.
(A function homogeneous in $Z$ of degree $p$ satisfies $f(kZ) = k^p f(Z)$,
so it has $U(1)$ charge $p$.)
For the scaling $\pi^a\rightarrow \kappa\pi^a,\,\bar\pi^a\rightarrow
\kappa^{-1}\bar
\pi^a$, each helicity component scales as $\kappa^{-2h}$ where $h$ is the
helicity of the state in Minkowski spacetime \cite{BW}. 
Thus $V_\phi(Z(\rho))$ describes the super gluon helicity states
$(1,{\bf 1}),({1\over 2},{\bf \bar 4}), (0,6), (-{1\over 2},{\bf 4}),
(-1,{\bf 1}).$ The spinor helicity variables $\pi^a, \bar\pi^{\dot a}$
are related to massless four-dimensional momentum $p_{a\dot a} = 
\pi_a\bar\pi_{\dot a} = \sigma_{a\dot a}^\mu p_\mu$ where
$\sigma^\mu = (1, \sigma^i)$ in terms of the Pauli matrices $\sigma^i$. 
Indices are raised and lowered $q_a = \epsilon_{ab}  q^b$, $q^a 
= \epsilon^{ab} q_b$,  $q_{\da} = \epsilon_{\da\db}  q^{\db}$, $q^{\da}
= \epsilon^{\da\db} q_{\db}$ with $\epsilon^{12}= 1 = -\epsilon_{12}$.
\vskip10pt 
%% F VERTICES
{\it $F$ Vertices}

For the conformal supergravity states, the vertex operator for the helicity 
states \break $(2,{\bf 1}), ({3\over 2},{\bf\bar 4}), (1,{\bf 6}), 
(\half, {\bf 4}), (0, {\bf 1})$ is given by 
\be
V_F(\rho) = f^{\dot a}(Z(\rho)) \,Y_{\dot a}(\rho)
\ee
with 
\begin{align}
f^{\da}(Z(\rho)) &= i \int \frac{dk}{k^2} \bpi^{\da} \prod^2_{a=1} 
\delta(k\lambda^a(\rho) -\pi^a) e^{ik\bpi_{\db}\mu^{\db}(\rho)} \cr
&\quad\times \left[ e_{2} + k\psi^b {\eta}_{\frac{3}{2}b} + 
\frac{k^2}{2} \psi^b\psi^c {T}_{1bc} + \frac{k^3}{3!} \psi^b\psi^c\psi^d 
{\Lambda}_{\frac{1}{2}bcd} + k^4 \psi^1\psi^2\psi^3\psi^4 \bar {C}_0 \right].
\end{align}
The function $f^{\dot a}(Z^I(\rho))$ is homogeneous in $Z^I(\rho)$ of degree 1.
The highest component (which is proportional to $e_2$) scales as 
$\kappa^{-4}$ with $\pi^a$ and $\bar\pi^a$, to describe helicity $2$. 
As required by the primary field conditions, 
${\partial\over\partial\mu^{\dot a}(\rho)} f^{\dot a}(Z(\rho)) = 0,$ 
since $\bar\pi_{\dot a} \bar\pi^{\dot a} = 0$. 
These vertices correspond to plane wave states and
diagonalize the translation generators. Together with the 
$F'$ vertices they comprise a dipole pair \cite{BW}. 

%% F' VERTICES
\vskip 20pt
{\it $F'$ Vertices}

The vertex operator for a second set of states
$(2,{\bf 1}), ({3\over 2},{\bf \bar 4}), (1,{\bf 6}), (\half, {\bf 4}), 
(0,{\bf 1})$ is 
\be
V_{F'}(\rho) = f^{a}(Z(\rho)) \,Y_{a}(\rho) 
+ \hat f^{\dot a}(Z(\rho)) \,Y_{\dot a}(\rho)
\ee
with

\begin{align}
f^a(Z(\rho)) &= \bs^a \int \frac{dk}{k^2} \prod^2_{a=1} 
\delta(k\lambda^a(\rho) - \pi^a) e^{ik\bpi_{\db}\mu^{\db}(\rho)} \cr
& \quad\times \left[ e_2' + k \psi^b {\eta}_{\frac{3}{2}b}'  
+ \frac{k^2}{2} \psi^b \psi^c {T}_{1bc}' + \frac{k^3}{3!} \psi^b \psi^c 
\psi^d {\Lambda}_{\half bcd}' + k^4\psi^1\psi^2\psi^3\psi^4 \bar{C}_0' 
\right]
\end{align}
and
\begin{align}
\hat{f}^{\da}(Z(\rho)) &= -is^{\da} \bs^e \int \frac{dk}{k^3} 
\frac{\partial}{\partial \lambda^e(\rho)} \prod^2_{a=1} 
\delta (k \lambda^a(\rho) - \pi^a) 
e^{ik\bpi_{\db}\mu^{\db}(\rho)} \cr
&\quad\quad\times \left[ e_2' + k \psi^b {\eta}_{\frac{3}{2}b}'  + 
\frac{k^2}{2} \psi^b \psi^c {T}_{1bc}' + \frac{k^3}{3!} \psi^b \psi^c 
\psi^d {\Lambda}_{\half bcd}' + k^4\psi^1\psi^2\psi^3\psi^4 \bar{C}_0' 
\right]\cr
\end{align}
chosen to satisfy the volume preserving condition 
${\partial\over \partial \lambda^a(\rho)}f^a(Z(\rho)) 
+ {\partial\over \partial \mu^{\dot a}(\rho)} \hat f^{\dot a}(Z(\rho)) = 0$.
The spinors  $s_{\dot a}$ and $\bar s_a$ are defined such that 
$\pi^a\bar s_{a} =1$ and $\bar\pi^{\dot a} s_{\dot a} = 1$.
These states are not eigenstates of the
momentum operator, as we discuss in (\ref{mo}).

%% G VERTICES
\vskip20pt 
{\it $G$ Vertices}

Conformal supergravity states with the 
opposite helicities and conjugate $SU(4)$ representations, 
$(0,{\bf 1}), (-{1\over 2},{\bf \bar 4}), (-1,{\bf 6}), (-{3\over 2} ,
{\bf 4}), (-2,{\bf 1})$ are described by
\be V_G(\rho) = g_a(Z(\rho)) \,\partial\lambda^a(\rho)
\ee
with
\begin{align}
g_a(Z(\rho)) &= \int dk\,k \,\lambda_a(\rho) \prod^2_{a=1} 
\delta(k \lambda^a(\rho) - \pi^a) e^{ik\bpi_{\db}\mu^{\db}(\rho)} \cr
&\quad\times \left[C_0 + k\psi^b\Lambda_{-\half b} + \frac{k^2}{2} 
\psi^b\psi^cT_{-1bc} + \frac{k^3}{3!} \psi^b\psi^c\psi^d\eta_{-\frac{3}{2}bcd} 
+k^4\psi^1\psi^2\psi^3\psi^4e_{-2} \right].\cr
\end{align}
$g_a(Z^J(\rho))$ is homogeneous in $Z^J(\rho)$ of degree $-1$.  
The highest component (proportional to $C$), scales with $\pi^a$ and  
$\bar \pi^a$ as $\kappa^0$ for zero helicity.   
Also, $\lambda^a(\rho) g_a(Z(\rho)) = 0$.
These are momentum eigenstates, and form a dipole pair with
the $G'$ vertices.
\vfill\eject

%% G' VERTICES

{\it $G'$ Vertices}

The final states that do not diagonalize the translation generators 
form a second set of states
$(0,{\bf 1}), (-{1\over 2},{\bf \bar 4}), (-1,{\bf 6}), (-{3\over 2} ,
{\bf 4}), (-2,{\bf 1})$ and correspond to 
\be
V_{G'}(\rho) = g_{\dot a}(Z(\rho)) \,\partial\mu^{\dot a}(\rho) + \hat g_{a}
(Z(\rho)) \,\partial\lambda^a(\rho)
\ee
with 
\begin{align}
g_{\da}(Z(\rho)) &= is_{\da} \int dk \prod^2_{a=1} 
\delta(k\lambda^a(\rho)-\pi^a) e^{ik\bpi_{\db}\mu^{\db}(\rho)} \cr
&\times \left[ C_0' + k\psi^b \Lambda_{-\half b}' + \frac{k^2}{2} 
\psi^b\psi^c T_{-1bc}' + \frac{k^3}{3!}\psi^b\psi^c\psi^d 
\eta_{-\frac{3}{2}bcd}' + k^4 \psi^1\psi^2\psi^3\psi^4 e_{-2}' \right]\cr
\end{align}
and
\begin{align}
\hat{g}_a(Z(\rho)) &= -i\bs_a s_{\da} \mu^{\da}(\rho) \int dk \, k 
\prod_{a=1}^2 \delta(k\lambda^a(\rho) - \pi^a) e^{ik\bpi_{\db}\mu^{\db}(\rho)} \cr
&\times \left[ C'_0 + k\psi^b \Lambda_{-\half b}' + \frac{k^2}{2} \psi^b\psi^c 
T_{-1bc}' + \frac{k^3}{3!}\psi^b\psi^c\psi^d \eta_{-\frac{3}{2}bcd}' + 
k^4 \psi^1\psi^2\psi^3\psi^4 e_{-2}' \right]\cr
\end{align}
with  $\mu^{\da}(\rho)g_{\da}(Z(\rho)) + \lambda^a(\rho) 
\hat g_a(Z(\rho)) =0$.

{\vskip20pt}
%% f VERTICES

{\it $f$ Vertices}

The vertex operator for the plane wave states with quantum numbers

$({\textstyle{3\over 2}}, {\bf 4}), \, (1, {\bf 15\oplus 1}), \,(
{\textstyle{1\over 2}}, 
\, {\bf \overline{20} \oplus \bar 4}), \,(0,{\bf 10\oplus 6}),\, 
(-{\textstyle{1\over 2}}, 
{\bf 4})$ is 
\be
V_f(\rho) = \;f^m\left(Z(\rho)\right) Y_m(\rho) \,\,+\,\, 
\tilde f^{\dot a} \left(Z(\rho)\right) Y_{\dot a}(\rho) 
\ee
with
\begin{align}
f^m(Z(\rho)) &= \int \frac{dk}{k^2} \prod^2_{a=1} \delta(k\lambda^a(\rho) 
- \pi^a) e^{ik\bpi_{\db} \mu^{\db}(\rho)} \cr
&\quad\times \left[ E^m_{\frac{3}{2}} + k\psi^b E^m_{1b} + \frac{k^2}{2} 
\psi^b\psi^c E^m_{\half bc} +\frac{k^3}{3!} \psi^b\psi^c\psi^d E^m_{0bcd} 
+ k^4 \psi^1\psi^2\psi^3\psi^4 E^m_{-\frac{1}{2}} \right]\cr
\end{align}
and
\begin{align}
\tilde{f}^{\da}(Z(\rho)) = -i s^{\da} &\int \frac{dk}{k^2} \prod^2_{a=1} 
\delta(k\lambda^a(\rho)-\pi^a) e^{ik\bpi_{\db}\mu^{\db}(\rho)} \cr
&\times \left[ E_{1m}^m + k\psi^c E^m_{\half mc} + \frac{k^2}{2} 
\psi^c\psi^d E^m_{0mcd}+ \frac{k^3}{3!} 
\psi^b\psi^c\psi^d \epsilon_{mbcd} E^m_{-\half} \right]\cr
\end{align}
so that ${\partial \over \partial \psi^m(\rho)} f^m(Z(\rho)) 
+ {\partial \over \partial \mu^{\dot a}(\rho)} \tilde f^{\dot a}(Z(\rho)) = 0$, 
$f^m(Z(\rho)$ and $\tilde f^{\dot a}(Z(\rho))$ have degree 1, and the leading 
components scale as $\kappa^{-3}$ and $\kappa^{-2}$ respectively. 

\vskip20pt
%%% g VERTICES

{\it $g$ Vertices}

The vertex operator for states with the opposite helicities and conjugate 
$SU(4)$ representations, $({\textstyle{1\over 2}}, {\bf \bar 4}), \, 
(0, {\bf \overline{10} \oplus 6}), \,({\textstyle-{1\over 2}}, \, 
{\bf 20 \oplus  4}), 
\,(-1,{ \bf 1\oplus 15}),\, (-{\textstyle{3\over 2}}, {\bf \bar 4})$ is 
\be
V_g(\rho) = g_m\left(Z(\rho)\right) \partial\psi^m(\rho)\,\,+ \,\, 
\tilde g_a\left(Z(\rho)\right) \partial\lambda^a(\rho)
\ee
with
\begin{align}
g_m(Z(\rho)) &= \int dk \prod^2_{a=1} \delta (k\lambda^a(\rho) -\pi^a) 
e^{ik\bpi_{\db}\mu^{\db}(\rho)} \cr
& \times \left[ \bar E_{\frac{1}{2}m} + k\psi^b\bar{E}_{0mb} + \frac{k^2}{2} 
\psi^b\psi^c \bar{E}_{-\half mbc} + \frac{k^3}{3!} \psi^b\psi^c\psi^d 
\bar{E}_{-1 mbcd} + k^4\psi^1\psi^2\psi^3\psi^4 \bar 
E_{-\frac{3}{2}m} \right]\cr
\end{align}
and
\begin{align}
\tilde{g}_a(Z(\rho)) = \bar{s}_a &\int dk \, k\prod^2_{a=1} \delta(k\lambda^a 
(\rho) -\pi^a) e^{ik\bpi_{\db}\mu^{\db}(\rho)} \cr
& \times \psi^m \left[ \bar E_{\frac{1}{2}m} + k\psi^b \bar{E}_{0mb} 
+ \frac{k^2}{2} 
\psi^b\psi^c\bar{E}_{-\half mbc} +\frac{k^3}{3!} \psi^b\psi^c\psi^d 
\bar{E}_{-1 mbcd} 
\right]\cr
\end{align}
where $\psi^m(\rho) g_m(Z(\rho)) \, + \, \lambda^a(\rho) \tilde g_a(Z(\rho)) 
= 0$.  
To obtain $\tilde g_a(Z(\rho))$, we use $\pi^a \bar s_a = 1$ which can be 
written as ${\pi1\over \lambda^1(\rho)}\lambda^a(\rho) \bar s_a = 1$ on the 
support of the delta function $\delta \left( {\lambda^2(\rho)\over 
\lambda^1(\rho)} - {\pi^2\over \pi^1}\right)$.

\subsection{\sl  Norm of the States}
We can check that the norms of the one particle supergraviton
states are zero.  
The physical states corresponding to the vertex operators are
shown in Table 2, 
%% TABLE
\begin{center}
\renewcommand{\tabcolsep}{1cm}
\renewcommand{\arraystretch}{1.25}
\vskip20pt
\begin{tabular}{| c | l |}
\hline
% {\bf Vertex Operator} & \hskip60pt {\bf State} \\ \hline
%
$V_F$ &  $f^{\da}(Z_0)Y_{\da(-1)}|0\rangle$  \\ \hline
$V_G$ & $g_a(Z_0) \lambda^a_{(-1)}|0\rangle$ \\ \hline
$V_{F'}$ & $\left(f^a(Z_0)Y_{a(-1)} + \hat{f}^{\da}(Z_0)Y_{\da(-1)}\right)|0\rangle$ \\ \hline
$V_{G'}$ & $\left(g_{\da}(Z_0) \mu^{\da}_{(-1)} + \hat{g}_a(Z_0)\lambda^a_{(-1)}\right)|0\rangle$ \\ \hline
$V_f$ & $\left(f^m(Z_0)Y_{m(-1)} + \tilde{f}^{\da}(Z_0)Y_{\da(-1)}\right)|0\rangle$ \\ \hline
$V_g$& $\left(g_m(Z_0)\psi^m_{(-1)} + \tilde{g}_a(Z_0)\lambda^a_{(-1)}\right)|0\rangle$ \\ \hline
$V_\Phi^A$ & $ V_{\phi}(Z_0)J^A_{(-1)}|0\rangle$ \\ \hline
\end{tabular}
\vskip10pt
\centerline{Table 2: One particle states}
\end{center}

where the  mode expansion for the twistor fields is 
$Z^I(\rho) = \sum_n Z^I_n\rho^{-n}, \; Y_J(\rho) = 
\sum_n Y_{Jn}\rho^{-n-1}, 
$
and the modes annihilating the vacuum are
$%\be
Z_n^I |0\rangle = 0, \; n\ge 1,$ and $Y_{nI} |0\rangle = 0, \;
n\ge 0.
$%\ee
\break The canonical commutation relations are
\be
\left[Z^I_n, Y_{Jm}\right] = \delta^I_J\,\delta_{n,-m}, 
\label{comrel}\ee 
and the hermitian conjugates \cite{DG1} are
$(Z_n^I)^\dagger = Z_{-n}^I$, for $1\le I \le 8$; and
$(Y_n^J)^\dagger = - Y_{-n}^J$, for $1\le J\le 4;$ and
$(Y_n^J)^\dagger = Y_{-n}^J$, for $5\le J\le 8$. 
We find that the only non-vanishing inner products for the
conformal graviton states are
$\langle 0 | V_{G'}^\dagger(0) V_F(0) |0\rangle$,
$\langle 0 | V_{F'}^\dagger(0) V_G(0) |0\rangle$,
and $\langle 0 | V_f^\dagger(0) V_g(0) |0\rangle$, so
the norms of the supergraviton states
vanish in the basis chosen in Table 2. 
We compute the inner products as follows. For example, 
\be
\langle 0| V_G^\dagger(0) V_F(0) |0\rangle 
= \langle 0| \lambda_1^a g_a^\ast(Z_0) f^{\dot a}(Z_0) Y_{\dot a
(-1)}|0\rangle = 0,
\ee
since $Z_0^I$ and $Y_{J(-1)}$ commute, and the $Y_{J(-1)}$ acting to the left
annihilate the vacuum.
%For $V_{F'}$,
%\begin{align}
%|| (f^a(Z_0)&Y_{a(-1)} + \hat f^{\da}(Z_0)Y_{\da(-1)})|0\rangle|| \cr
%&= \langle 0|(Y_{\db(1)} (\hat f^{\db}(Z_0))^\ast 
%+ Y_{b(1)}(f^b(Z_0)^\ast)(f^a(Z_0)Y_{a(-1)} 
%+ \hat f^{\da}(Z_0)Y_{\da(-1)})|0\rangle = 0.\cr
%\end{align}
%Similarly, the norms for the states $V_G,V_{G'}, V_f,V_g$ all vanish,
%since they involve different $Y$ and $Z$ modes and thus commute,
%allowing the negative $Y$ modes to annihilate the left vacuum. 
In contrast, the gluon norm is positive,
\begin{align}
& || V_{\phi}(Z_0) J^A_{-1}|0\rangle || = \langle 0|J^A_{1}V^\ast_{\phi}(Z_0) 
V_{\phi}(Z_0) J^A_{-1}|0\rangle \cr
&= \langle 0| J^A_{1}J^A_{-1}|0\rangle \int dZ_0 |V_{\phi}(Z_0)|^2 
=k \int dZ_0 |V_{\phi}(Z_0)|^2 >0, \end{align}
where $k$ is the level of the current algebra,
$J^A_n J^B_m = if^{AB}_{\hskip12pt C} J^C_{n+m} +  k n \delta_{n,-m}\delta^{AB}
$.
\vskip-20pt
\subsection{\sl A Subset of Vertex Operators}
\vskip-5pt
We will focus on amplitudes involving a subset of the vertex operators,
relabeled in Table 3.
\vskip-20pt
\begin{center}
\renewcommand{\arraystretch}{1.75}
\begin{tabular}{c | r c l |} \hline
\multicolumn{1}{|c|}{\multirow{2}{*}{$V_F$}} & $e_2(\rho)$ & $=$ & $i \int dk \,k^{-2} \prod^2_{a=1} \delta(k\lambda^a(\rho) -\pi^a) e^{ik\bpi_{\db}\mu^{\db}(\rho)} \, \bpi^{\da} Y_{\da}(\rho) \,e_{2}$ \\ \cline{2-4}
\multicolumn{1}{|c|}{} & $\bar C(\rho)$ &$=$&$ i \int dk \,k^{-2} \prod^2_{a=1} \delta(k\lambda^a(\rho) -\pi^a) e^{ik\bpi_{\db}\mu^{\db}(\rho)} \, k^4 \psi^1\psi^2\psi^3\psi^4 \, \bpi^{\da} Y_{\da}(\rho)\,\bar C_0$ \\ \hline
\multicolumn{1}{|c|}{\multirow{2}{*}{$V_G$}} & $C(\rho)$ &$=$&$ \int dk\,k  
\prod^2_{a=1} \delta(k \lambda^a(\rho) - \pi^a) e^{ik\bpi_{\db}\mu^{\db}(\rho)
} \,\lambda_a(\rho)\partial\lambda^a(\rho) \, C_0$ \\ \cline{2-4}
\multicolumn{1}{|c|}{} & $e_{-2}(\rho)$ &$=$&$ \int dk\,k \prod^2_{a=1} 
\delta(k \lambda^a(\rho) - \pi^a) e^{ik\bpi_{\db}\mu^{\db}(\rho)} \,
k^4\psi^1\psi^2\psi^3\psi^4 \, \lambda_a(\rho) \partial \lambda^a(\rho)\, 
e_{-2}$ \\ \hline
\multicolumn{1}{|c|}{\multirow{4}{*}{$V_{F'}$}} & $e'_2(\rho)$ & $=$&$ 
\int dk\,k^{-2} \hskip-3pt \left [ \bs^a Y_a(\rho)\,
\prod^2_{a=1} \delta(k\lambda^a(\rho) 
- \pi^a) \, \right. $ \\
\multicolumn{1}{|c|}{} & & & $ \left. \qquad
\qquad + i\bar s^b \big({\partial\over\partial\pi^b} \prod^2_{a=1} \delta(k\lambda^a(\rho) - \pi^a)\big) s^{\dot a}Y_{\dot a}(\rho)\right] e^{ik\bpi_{\db}\mu^{\db}(\rho)}\nobreak e'_2$ \\ \cline{2-4}
\multicolumn{1}{|c|}{} & $\bar C'(\rho)$ & $=$&$\int dk\,k^{-2} \hskip-3pt 
\left [ \bs^a Y_a(\rho)\,\prod^2_{a=1} \delta(k\lambda^a(\rho) - \pi^a) 
\,\right. $ \\
\multicolumn{1}{|c|}{} & & & $\qquad
\qquad \left. +i\bar s^b \big({\partial\over
\partial\pi^b} \prod^2_{a=1} \delta(k\lambda^a(\rho) - \pi^a)\big) s^{\dot a}
Y_{\dot a}(\rho)\right] e^{ik\bpi_{\db}\mu^{\db}(\rho)} k^4 \psi^1\psi^2
\psi^3\psi^4 \,\bar C_0'$ \\ \hline
\multicolumn{1}{|c|}{\multirow{3}{*}{$V_{G'}$}} & 
$C'(\rho) $&$=$&$  i \int dk\,k\prod^2_{a=1} \delta(k\lambda^a(\rho) -\pi^a)
\left [ k^{-1} s_{\da} \partial \mu^{\da}(\rho) - s_{\da} \mu^{\da}(\rho) 
\bar s_a \partial \lambda^a(\rho)\right] e^{ik\bpi_{\db}\mu^{\db}(\rho)} C'_0$
 \\ \cline{2-4}
\multicolumn{1}{|c|}{} & $e'_{-2}(\rho) $&$=$&$ i \int dk\,k\prod^2_{a=1} 
\delta(k\lambda^a(\rho) -\pi^a) \left [ k^{-1} s_{\da} \partial \mu^{\da}(\rho) 
- s_{\da} \mu^{\da}(\rho) \bar s_a \partial \lambda^a(\rho)\right]
$\\
\multicolumn{1}{|c|}{} & & & $ \hskip150pt
\times \,%- s_{\da} \mu^{\da}(\rho) \bar s_a \partial \lambda^a(\rho)\right] 
e^{ik\bpi_{\db}\mu^{\db}(\rho)}
%+ i\bar s^b 
%\big({\partial\over\partial\pi^b} \prod^2_{a=1} \delta(k\lambda^a(\rho) 
%- \pi^a)\big) s^{\dot a}Y_{\dot a}(\rho)\right] e^{ik\bpi_{\db}\mu^{\db}(\rho)}
%\hskip-3pt:\nobreak e'_2$ \\ \hline
\,k^4 \psi^1\psi^2\psi^3\psi^4 \, e'_{-2}$ \\ \hline
\multicolumn{1}{|c|}{\multirow{2}{*}{$V_{\Phi}$}} & $A^A_1(\rho) $&$=$&$ \int dk \,k^{-1} \prod^2_{a=1}\delta(k\lambda^a(\rho) -\pi^a) e^{ik\bar\pi_{\dot b} \mu^{\dot b}(\rho)} \,A_1\ J^A(\rho)$ \\ \cline{2-4}
\multicolumn{1}{|c|}{} & $A^A_{-1}(\rho) $&$=$&$ \int dk \,k^{-1} \prod^2_{a=1}\delta(k\lambda^a(\rho)
-\pi^a) e^{ik\bar\pi_{\dot b} \mu^{\dot b}(\rho)} k^4 \psi^1\psi^2\psi^3\psi^4
\,A_{-1}\, J^A(\rho)$ \\ \hline
\end{tabular}
\vskip1pt
\centerline{Table 3: A subset of the vertex operators: for conformal gravitons, 
scalars and gluons}
\end{center}
\vfill\eject

% SECTION
\section{Three-Point Couplings}

In this section, 
we compute the non-vanishing three-point amplitudes for the
gravitons, scalars, and gluons  in the $V_F$, $V_G$ and $V_\Phi$ vertices
using canonical quantization, and then extend these to
the corresponding states in the primed vertices, $V_{F'},$ and $V_{G'}$.
We have relabeled this subset of vertex operators for positive and negative
helicity states in Table 3. (It will be convenient to consider the scalars
$\bar C,\bar C'$ as negative helicity, and $C,C'$ as positive helicity
when computing amplitudes, as in \cite{BW}.) Amplitudes for  
other states can be calculated with similar ease.

Scattering amplitudes in twistor string theory receive contributions from
the various instanton sectors, which are due to world sheet gauge fields
\cite{W, B}.
Amplitudes with the number of negative helicity states equal to
$d+1-\ell$, are computed with instanton number $d$, where $\ell$ is 
the number of loops. For tree amplitudes, $\ell = 0$. 
We compute the $N$-point tree as \cite{DG1}
\be
\langle V_1(\rho_1) V_2(\rho_2)\ldots V_N(\rho_N) \rangle_{\rm tree}
= \int 
\langle 0| e^{dq_0}V_1(\rho_1)V_2(\rho_2) \cdots V_N(\rho_N) |0\rangle 
\prod_{r=1}^N d\rho_r / d\gamma_Md\gamma_S\ee
where $d\gamma_M$ is the invariant measure of the Mobius group, and
$d\gamma_S$ is the invariant measure of the scaling group.
$q_0$ is the conjugate zero mode of the $U(1)$ current
and commutes with field modes as $Y^I_{n-d} e^{dq_0} =  e^{dq_0} Y_n^I$ and
$Z_{n+d}^I e^{dq_0} = e^{dq_0} Z_n^I$.

\subsection{\sl Unprimed Couplings}
Using the canonical methods of \cite{DG1},
we compute the non-vanishing three-point tree amplitudes
that come from the degree one curves as follows.
\begin{align}
\langle A_{-1}^{A_1}&(\rho_1)A_{-1}^{A_2}(\rho_2)C(\rho_3) \rangle_{\rm tree} 
= \int \langle 0| e^{q_0} A_{-1}^{A_1}(\rho_1)A_{-1}^{A_2}(\rho_2)C(\rho_3) 
|0 \rangle \prod_{r=1}^3 d\rho_r / d\gamma_Md\gamma_S\cr
&=\int \prod^3_{r=1}dk_r k_r \lambda_a(\rho_3) \partial 
\lambda^a(\rho_3) \prod_{r,a} \delta (\pi_r{}^a - k_r\lambda^a(\rho_r)) 
(\rho_1 -\rho_2)^4(k_1k_2)^4 \cr
&\quad\times \langle 0| e^{q_0} e^{i\sum_{r=1}^3k_r\bpi_{r\db}\mu^{\db}
(\rho_r)} |0\rangle \prod_a d^2\lambda^a \prod_r d\rho_r/ d\gamma_M d\gamma_S 
\left( {-\delta^{A_1A_2}A_{-1(1)}A_{-1(2)}C_{0(3)}\over (\rho_1 -\rho_2)^2
k_1^2k_2^2}\right) \cr
&=- \delta^4(\Sigma\pi_r\bpi_r) \int \prod_{r=1}^3d\zeta_r \prod_{r=1}^3\pi_r{}^1
\delta(\pi_r{}^2 - \zeta_r \pi_r{}^1) (\zeta_1 - \zeta_2)^2(\pi_1{}^1
\pi_2{}^1)^2\,\delta^{A_1A_2}A_{-1(1)}A_{-1(2)}C_{0(3)}
\cr
&=-\delta^4(\Sigma\pi_r\bpi_r)\langle 12 \rangle^2 \delta^{A_1A_2}
A_{-1(1)}A_{-1(2)}C_{0(3)}
= \epsilon^-_1 \cdot p_2 \, \epsilon^-_2\cdot p_1 \, 
\delta^4(\Sigma\pi_r\bpi_r)\delta^{A_1A_2} C_{0(3)}.\cr
\label{ggc}\end{align}
This amplitude is type $\phi\phi G$.
We replace $Z^I(\rho)$ with $Z_0^I + \rho Z_{-1}^I$, as the
only surviving modes, and
change variables $\zeta_r = {\lambda^2(\rho_r)\over \lambda^1(\rho_r)}$.
For $d=1$, the invariant measures are expressed as
$d\gamma_Md\gamma_S=
\prod_{a=1}^2 d^2\lambda^a (\det \lambda)^{-2}$ where
$\prod_a d^2\lambda^a = d\lambda_0^1d\lambda_{-1}^1
d\lambda^2_0d\lambda^2_{-1},$
and $\det \lambda = \lambda^1_0\lambda^2_{-1} - \lambda^2_0\lambda^1_{-1}$.
The current algebra contribution follows from (\ref{fzh}).
The gluon polarizations are given by  
$\epsilon^-_r=A_{-1(r)}\pi_{ra} s_{r\dot a}$ and
$\epsilon^+_r=A_{1(r)}\bar s_{ra}\bar\pi_{r\dot a}$ . 
We use momentum conservation and thus $s_{s\dot b} \sum_r \pi_r^b\bar
\pi_r^{\dot b}= 0$ to find $s_{1\dot b} \bar\pi_2^{\dot b} 
= {\langle 31\rangle\over \langle 23\rangle}$ and
$s_{2\dot b} \bar\pi_1^{\dot b} = {\langle 23\rangle\over \langle 31\rangle}$, 
so that $\epsilon_1^-\cdot p_2\,\epsilon_2^-\cdot p_1 = -\langle 12\rangle^2 \; 
A_{-1(1)} A_{-1(2)}$, with $\langle rs\rangle = \pi_{ra}\pi_s^a$
and $[rs] = \bar\pi_{r\dot a}\bar\pi^{\dot a}$.
We can set the scalar wave function $C_{0(3)} =1$.
The momentum conserving delta functions are
\be \prod_{\da,b}\delta\left(\Sigma_{r=1}^3 \pi_r^b\bpi_{r\da}\right)
\equiv \delta^4(\Sigma\pi_r\bpi_r).\ee
\vskip20pt 

For two gluons and a graviton ($\phi\phi G$), 
\begin{align}
&\langle
A_{1}^{A_1}(\rho_1) A_{-1}^{A_2}(\rho_2)
e_{-2}(\rho_3) \rangle_{\rm tree} = \int \langle 0|
e^{q_0} A_1^{A_1}(\rho_1) A_{-1}^{A_2}(\rho_2)e_{-2}(\rho_3) | 0\rangle
\prod_{r=1}^3 d\rho_r / d\gamma_Md\gamma_S\cr
&= -\int \prod_{r=1}^3 dk_r \prod_{r=1}^3 k_r \;\lambda_a(\rho_3)\partial
\lambda^a(\rho_3)\, \prod_{r,a}\delta(\pi_r{}^a-k_r\lambda^a(\rho_r)) \; 
(\rho_2-\rho_3)^4(k_2k_3)^4\cr
&\hskip40pt \times\langle 0|e^{q_0} e^{\sum_{r=1}^3 ik_r\bar\pi_{r\db}\mu^{\db}
(\rho_r)}|0\rangle \prod_a d^2\lambda^a\prod_rd\rho_r/d\gamma_Sd\gamma_M 
\left({\delta^{A_1 A_2} A_{1(1)} A_{-1(2)} e_{-2(3)}\over (\rho_1-\rho_2)^2
(k_1k_2)^2}\right)\cr
&=-\delta^4(\Sigma\pi_r\bar\pi_r) \; \int \prod_{r=1}^3 d\zeta_r \, 
\prod_{r=1}^3 \pi_r^1\, \delta(\pi^2_r - \zeta_r \pi_r^1)(\zeta_2-\zeta_3)^4 
(\pi_2^1\pi_3^1)^4 (\zeta_1-\zeta_2)^{-2} (\pi_1^1\pi_2^1)^{-2}\cr
&\hskip40pt\times
\delta^{A_1 A_2} A_{1(1)} A_{-1(2)} 
e_{-2(3)}\cr
&= -\delta^4(\Sigma\pi_r\bar\pi_r)\, \frac{\langle 23\rangle^4}
{\langle 12\rangle^2} \; \delta^{A_1 A_2} A_{1(1)} A_{-1(2)} e_{-2(3)}\cr
&= \left (\epsilon_1^+\cdot \epsilon_2^-\;\epsilon_{3\,{a\dot a b\dot b}}^- 
p_1^{a\dot a} p_2^{b\dot b} +\epsilon_1^+\cdot p_2 \;\epsilon_{3\,a\dot a b
\dot b}^- \epsilon_2^{- a\dot a} p_2^{b\dot b} + \epsilon_2^-\cdot p_3 \;
\epsilon_{3\,a \dot a b\dot b}^-  \epsilon_1^{+\, a\dot a} p_2^{b\dot b}\right) 
\delta^{A_1 A_2}\, \delta^4(\Sigma\pi_r\bar\pi_r).\cr
\label{gge}\end{align}
The gravity polarizations are 
$\epsilon^-_r =e_{-2(r)}\pi_{ra} s_{r\dot a}\pi_{rb} s_{r\dot b}$ 
and $\epsilon^+_r=e_{2(r)}\bar s_{ra}\bar\pi_{r\dot a} \bar s_{rb}
\bar\pi_{r\dot b}$, and one can factor 
\begin{align}
&\epsilon_1^+\cdot \epsilon_2^-\;\epsilon_{3\,{a\dot a b\dot b}}^-p_1^{a\dot a} 
p_2^{b\dot b} +\epsilon_1^+\cdot p_2 \;\epsilon_{3\,a\dot a b\dot b}^- 
\epsilon_2^{- a\dot a} p_2^{b\dot b} + \epsilon_2^-\cdot p_3 \;
\epsilon_{3\,a \dot a b\dot b}^-  \epsilon_1^{+\, a\dot a} p_2^{b\dot b}\cr
& = \left (\epsilon_1^+\cdot \epsilon_2^-\;\epsilon_3^- \cdot p_1 
+\epsilon_1^+\cdot p_2 \;\epsilon_3^- \cdot \epsilon_2^- + 
\epsilon_2^-\cdot p_3 \;\epsilon_3^- \cdot \epsilon_1^+ \right ) 
\epsilon_3^- \cdot p_2 
= {\langle 23\rangle^3\over \langle 12\rangle \langle 31\rangle}\, 
{\langle 23\rangle \langle 31\rangle\over \langle 12\rangle} 
= {\langle 23\rangle^4\over \langle 12\rangle ^2}.\cr
\end{align}

For two gravitons and a scalar ($GGG$),
\begin{align}
&\langle e_{-2}(\rho_1) e_{-2}(\rho_2) C(\rho_3) \rangle_{\rm tree} = 
\int \langle 0| e^{q_0} e_{-2}(\rho_1) e_{-2}(\rho_2)C(\rho_3) | 0\rangle
\prod_{r=1}^3 d\rho_r / d\gamma_Md\gamma_S\cr
&= \int \prod_{r=1}^3 dk_r \prod_{r=1}^3 k_r \; \lambda_a(\rho_1)\partial
\lambda^a(\rho_1)\, \lambda_b(\rho_2)\partial\lambda^b(\rho_2)\, 
\lambda_c(\rho_3)\partial\lambda^c(\rho_3)\, 
\prod_{ra}\delta(\pi^a_r-k_r\lambda^a(\rho_r)) \; (\rho_1-\rho_2)^4k_1^4k_2^4\cr
&\hskip40pt \times\langle 0|e^{q_0} e^{\sum_{r=1}^3 ik_r\bar\pi_{r\dot b}
\mu^{\dot b}(\rho_r)}|0\rangle \prod_a d^2\lambda^a\prod_rd\rho_r/d\gamma_S
d\gamma_M\, e_{-2(1)} e_{-2(2)} C_{0(3)}\cr
&= \delta^4(\Sigma\pi_r\bar\pi_r) \; \int \prod_{r=1}^3 d\zeta_r \, 
\prod_{r=1}^3 \pi_r^1\, \delta(\pi^2_r - \zeta_r \pi_r^1)\, 
(\zeta_1-\zeta_2)^2 (\pi_1^1\pi_2^1)^2 \, e_{-2(1)} e_{-2(2)} C_{0(3)}\cr
&= \delta^4(\Sigma\pi_r\bar\pi_r)\, \langle 12\rangle^4 e_{-2(1)} e_{-2(2)} 
C_{0(3)} 
= \epsilon_{1\, a\dot a b\dot b}^- p_2^{a\dot a} p_2^{b\dot b} \,
\epsilon_{2\, c\dot c d\dot d}^- p_1^{c\dot c} p_1^{d\dot d}\, \;
\delta^4(\Sigma\pi_r\bar\pi_r)\,\;C_{0(3)}.\label{eec}
\end{align}
\vskip10pt

Less conventional is the three-graviton coupling ($GGF$):
\begin{align}
&\langle e_{-2}(\rho_1) e_{-2}(\rho_2) e_2(\rho_3) \rangle_{\rm tree} 
= \int \langle 0| e^{q_0} e_{-2}(\rho_1) e_{-2}(\rho_2)e_2(\rho_3) 
| 0\rangle\prod_{r=1}^3 d\rho_r / d\gamma_Md\gamma_S\cr
&= i \int \prod_{r=1}^3 dk_r  {k_1 k_2\over k_3^2}\; \lambda_a(\rho_1)
\partial\lambda^a(\rho_1)\, \lambda_a(\rho_2)\partial\lambda^a(\rho_2)\, 
\prod_{ra}\delta(\pi^a_r-k_r\lambda^a(\rho_r)) \; (\rho_1-\rho_2)^4k_1^4k_2^4\cr
&\hskip40pt \times\langle 0|e^{q_0} e^{\sum_{r=1}^3 ik_r\bar\pi_{r\db}
\mu^{\db}(\rho_r)}\, \bar\pi_3^{\dot a} Y_{\dot a}(\rho_3)|0\rangle \prod_a d^2
\lambda^a\prod_rd\rho_r/d\gamma_Sd\gamma_M e_{-2(1)} e_{-2(2)} e_{2(3)}\cr
&= i \delta^4(\Sigma\pi_r\bar\pi_r) \; \int \prod_{r=1}^3 d\zeta_r \, 
\prod_{r=1}^3 \, \delta(\pi^2_r - \zeta_r \pi_r^1)\, (\zeta_1-\zeta_2)^{4} 
(\pi_1^1\pi_2^1)^{4} \, {\pi_1^1\pi_2^1\over(\pi_{3}^1)^2}\; (-i) \,
\sum_{r=1}^2 {\pi_r^1 [3r]\over \zeta_r-\zeta_3}\; \;e_{-2(1)} e_{-2(2)} 
e_{2(3)}\cr
&= \delta^4(\Sigma \pi_r\bar\pi_r)\, \langle 12\rangle^4\; \,\sum_{r=1}^2 
{[3r]\langle r\xi\rangle^2\over \langle 3r\rangle \langle 3\xi\rangle^2} 
\,e_{-2(1)} e_{-2(2)} e_{2(3)}
\cr&= \delta^4(\Sigma \pi_r\bar\pi_r)\, \langle 12\rangle^6\; {[32]\over 
\langle 32\rangle \langle 31\rangle^2} \, e_{-2(1)} e_{-2(2)} e_{2(3)} = 0,
\label{eee}
\end{align}
since $\langle 12\rangle [23] = 0$ by momentum conservation.  
Since this amplitude involves $Y_{\dot a}$, 
we have first evaluated, using (\ref{comrel}), 
\be 
\langle 0|e^{q_0} e^{\sum_{r=1}^3 ik_r\bar\pi_{r\db}\mu^{\db}(\rho_r)}\, 
\bar\pi_3^{\dot a} Y_{\dot a}(\rho_3)|0\rangle 
= -i\sum_{r\ne 3} {k_r[3r]\over(\rho_r-\rho_3)}
\langle 0|e^{q_0} e^{\sum_{r=1}^3 ik_r\bar\pi_{r\db}\mu^{\db}(\rho_r)}|0\rangle 
,\label{yvac}\ee
then replaced $\mu^{\db}(\rho)$ by its lowest modes and changed variables
from $\rho_r$ to $\zeta_r$, as discussed in more detail in (\ref{chv}).
The expression is independent of the spinor $\xi$.
We compare this vanishing three-graviton tree amplitude for conformal gravity 
with that of Einstein gravity, 
\begin{align}
\langle e_{-2}(\rho_1) e_{-2}(\rho_2) e_2(\rho_3) \rangle_{
Einstein\,  tree}  
&= {\langle 12\rangle^6\; \over \langle 23\rangle^2 \langle 31\rangle^2} \;  
\delta^4(\Sigma\pi_r\bpi_r)\, e_{-2(1)} e_{-2(2)} e_{2(3)} \ne 0\cr & 
= {1\over s_{23}} \langle e_{-2}(\rho_1) e_{-2}(\rho_2) e_2(\rho_3) 
\rangle_{\rm tree}.
\end{align}
For two scalars and a graviton ($GGF$), the amplitude also vanishes
by momentum conservation:
\begin{align}
\langle &C(\rho_1) e_{-2}(\rho_2) \bar C(\rho_3) \rangle_{\rm tree} = 
\int \langle 0| e^{q_0} C(\rho_1) e_{-2}(\rho_2)\bar C(\rho_3) | 0\rangle 
\prod_{r=1}^3 d\rho_r / d\gamma_Md\gamma_S\cr
&=  i \int \prod_{r=1}^3 dk_r  {k_1 k_2\over k_3^2}\; \lambda_a(\rho_1)
\partial\lambda^a(\rho_1)\, \lambda_a(\rho_2)\partial\lambda^a(\rho_2)\, 
\prod_{ra}\delta(\pi^a_r-k_r\lambda^a(\rho_r)) \; (\rho_2-\rho_3)^4k_2^4k_3^4\cr
&\hskip40pt \times\langle 0|e^{q_0} e^{\sum_{r=1}^3 ik_r\bar\pi_{r\db}
\mu^{\db}(\rho_r)}\, \bar\pi_3^{\dot a} Y_{\dot a}(\rho_3)|0\rangle \prod_a d^2
\lambda^a\prod_rd\rho_r/d\gamma_Sd\gamma_M\quad C_{0(1)} e_{-2(2)} 
\bar C_{0(3)}\cr
&= \delta^4(\Sigma \pi_r\bpi_r)\, \langle 12\rangle^2\; 
{[23]\langle 23\rangle^3\over \langle 31\rangle^2} C_{0(1)} e_{-2(2)} 
\bar C_{0(3)} = 0.
\end{align}
The remaining three-point functions with two negative helicity states
also vanish.
For comparison, we include the familiar degree one three-point gluon vertex,
\begin{align}
\langle A_{-1}^{A_1}&(\rho_1) A^{A_2}_{-1}(\rho_2) A^{A_3}_1(\rho_3) 
\rangle_{\rm tree} = \int \langle 0| e^{q_0} A_{-1}^{A_1}(\rho_1) 
A_{-1}^{A_2}(\rho_2)A_1^{A_3}(\rho_3) | 0\rangle\prod_{r=1}^3 d\rho_r / 
d\gamma_Md\gamma_S\cr
&= \int \prod_{r=1}^3 {dk_r \over k_r} \prod_{ra}\delta(\pi^a_r-k_r
\lambda^a(\rho_r)) \; (\rho_1-\rho_2)^4k_1^4k_2^4\; 
{f^{A_1A_2 A_3}\over (\rho_1-\rho_2)(\rho_2-\rho_3)(\rho_3-\rho_1)}\cr
&\hskip40pt \times\langle 0|e^{q_0} e^{\sum_{r=1}^3 
ik_r\bar\pi_{r\dot b}\mu^{\dot b}(\rho_r)}|0\rangle \prod_a d^2\lambda^a\prod_r
d\rho_r/d\gamma_Sd\gamma_M\,A_{-1(1)} A_{-1(2)} A_{1(3)}\cr
%&= \delta^4(\Sigma \pi^a\bar\pi_b) \; \int \prod_{r=1}^3 d\zeta_r \, \prod_{r=1}3 \pi_r1\, \delta(\pi_r{}^2 - \zeta_r \pi_r{}^1)\,(\zeta_1-\zeta_2)^4 (\pi_11\pi_21)^4 \, \cr
%&\hskip40pt \times \prod_{r=1}^3 {1\over (\pi_r^1)^2}\;{A_{-1(1)} A_{-1(2)} A_{1(3)} f^{a_1a_2a_3}\over (\zeta_1- \zeta_2)(\zeta_2-\zeta_3)(\zeta_3-\zeta_1)}\cr 
%&= \delta^4(\Sigma \pi^a\bar\pi_b)\, {\langle 12\rangle^4\over \langle 12\rangle \langle 23\rangle \langle 31\rangle} \;  f^{a_1a_2a_3}\; A_{-1(1)} A_{-1(2)} A_{1(3)} \cr
&= \delta^4(\Sigma \pi_r\bpi_r)\, {\langle 12\rangle^3\over \langle 23
\rangle \langle 31\rangle}\; f^{A_1A_2A_3}\; A_{-1(1)} A_{-1(2)} A_{1(3)}\cr 
&= \delta^4(\Sigma \pi_r\bar\pi_r)\;f^{A_1A_2A_3}\;
\left (\epsilon_1^-\cdot \epsilon_2^- 
\epsilon_3^+\cdot p_1 + \epsilon_2^-\cdot \epsilon_3^+ \epsilon_1^-\cdot p_2 
+\epsilon_3^+\cdot \epsilon_1^- \epsilon_2^-\cdot p_3\right).
\end{align}
The unprimed MHV three-point functions are summarized in Table 4, where
we include their polarizations and momentum conserving delta function, in
order to compare with primed couplings in Table 6. 
Our calculations agree with the general Berkovits-Witten formula \cite{BW},
derived from path integral methods, and our analysis 
is useful in extending to the dipole states, and in comparing
with conventional field theory couplings.  
 
\begin{center}
\renewcommand{\tabcolsep}{1cm}
\renewcommand{\arraystretch}{1.7}
\vskip20pt
\begin{tabular}{| l |} \hline
%{\bf Unprimed MHV Couplings} \\ \hline
$\langle A^{A_1}_{-1}A^{A_2}_{-1}C\rangle = -
\langle 12 \rangle^2\,\delta^{A_1A_2} A_{-1(1)}A_{-1(2)}C_{0(3)}
\delta^4(\Sigma \pi_r \bpi_r)$
\\ \hline
$\langle A_1^{A_1}A_{-1}^{A_2}e_{-2} \rangle = -
{\langle 23 \rangle^4\over \langle 12 \rangle^2} \delta^{A_1A_2}
A_{1(1)}A_{-1(2)}e_{-2(3)}\delta^4(\Sigma \pi_r \bpi_r)$\\ \hline
$\langle e_{-2}e_{-2}C\rangle = 
\langle 12 \rangle^4  e_{-2(1)}e_{-2(2)}C_{0(3)}
\delta^4(\Sigma \pi_r \bpi_r)$\\ \hline
$\langle e_{-2}e_{-2}e_2 \rangle = 
{\langle 12\rangle^6 [23]\over \langle 23 \rangle \langle 31\rangle^2} 
e_{-2(1)}e_{-2(2)}e_{2(3)}\delta^4(\Sigma \pi_r \bpi_r)=0$\\ \hline
$\langle Ce_{-2} \bar C \rangle = 
{\langle 12 \rangle^2 \langle 23 \rangle^3 [23]\over \langle 31 \rangle^2} 
C_{0(1)}e_{-2(2)}\bar C_{0(3)}\delta^4(\Sigma \pi_r \bpi_r)= 0$\\ \hline
$\langle A^{A_1}_{-1}A^{A_2}_{-1}A^{A_3}_1 \rangle 
= {\langle 12 \rangle^3 \over \langle 23\rangle 
\langle 31\rangle} f^{A_1 A_2 A_3}A_{-1(1)}A_{-1(2)}A_{1(3)}
 \delta^4(\Sigma \pi_r \bpi_r)$\\ \hline
\end{tabular}
\vskip10pt
\centerline{Table 4: Unprimed conformal supergravity MHV couplings}
\end{center}

We compare these couplings with those of opposite helicities, with 
instanton number zero:
\begin{align} \langle A_{1}^{A_1}&(\rho_1) A_{1}^{A_2}(\rho_2) 
\bar C(\rho_3) \rangle_{\rm tree} = \int \langle 0| A_1^{A_1}(\rho_1) 
A_1^{A_2}(\rho_2)\bar C(\rho_3) | 0\rangle\prod_{r=1}^3 d\rho_r / 
d\gamma_Md\gamma_S\cr 
&=  i  \int \prod_{r=1}^3 dk_r {k_3^2\over k_1k_2} \; \prod_{r,a}
\delta(\pi_r{}^a-k_r\lambda^a) 
\langle 0| e^{\sum_{r=1}^3 ik_r\bar\pi_{r\db}
\mu^{\db}(\rho_r)}\;\bar\pi^{\dot a}_3 Y_{\dot a}(\rho_3)\,|0\rangle\cr
&\hskip40pt \times \prod_a d\lambda^a \prod_rd\rho_r/d\gamma_Sd\gamma_M\, 
\left( -{\delta^{A_1 A_2} A_{1(1)} A_{1(2)} \bar C_{0(3)} \over(\rho_1-\rho_2)^2
}\right)\cr
&= -\int \prod_{r=1}^3 dk_r {k_3^2\over k_1k_2} \; \prod_{ra}
\delta(\pi^a_r-k_r\lambda^a)\;  \left( \sum_{r\ne 3} {k_r [3r]\over 
(\rho_r-\rho_3)} \right) \; \prod_{a=1}^2 \delta(\sum_{r=1}^3 
k_r \bar\pi_{ra})\cr
&\hskip40pt \times \prod_a d\lambda^a \prod_rd\rho_r/d\gamma_Sd\gamma_M
\quad \delta^{a_1 a_2} A_{1(1)}A_{1(2)} \bar C_{0(3)} 
\,(\rho_1-\rho_2)^{-2}\cr
&= - \delta^{a_1a_2} [31] {(\pi_3^1)^2\over \pi_2^1} \int {d\lambda^2\over
\lambda^1}\prod_{r=1}^3\delta(\pi_r^2 - {\lambda^2\over \lambda^1} \pi_r^1)\;
\prod_a\delta(\sum_{r=1}^3 {\pi_r^1\over\lambda^1}
\bar\pi_{ra})\;A_{1(1)} A_{1(2)} 
\bar C_{0(3)}\cr
%&= - \delta^{a_1a_2}\;\delta^4(\Sigma \pi^a\bar\pi_b) \;  {(\pi_3^1)^2\over \pi_1^1\pi_2^1} [23] [31] \, A_{1(1)} A_{1(2)} \bar C_{0(3)}\cr
%%
&= - \,\delta^{A_1A_2}\;\delta^4(\Sigma \pi_r\bar\pi_r) \;  [12]^2 \, 
A_{1(1)} A_{1(2)} \bar C_{0(3)} .
\label{ggbarc}\end{align}
After eliminating $Y_{\db}$, for degree $d=0$ 
we replace $Z^I(\rho)$ with $Z_0^I$. From momentum conservation,  
we find $\pi_2^1[21] = -\pi_3^1 [31]$. We can replace two of the delta
functions $\prod_{r=2}^3 
\delta(\pi_r^2 - {\lambda^2\over \lambda^1} \pi_r^1)$ with 
$\prod_{a=1}^2 \delta(\sum_{r=1}^3 (\pi_r^2 - {\lambda^2 \over \lambda^1} 
\pi_r^1)\bar\pi_{ra}) \, [23] = \prod_{a=1}^2 \delta(\sum_{r=1}^3 
\pi_r^2\bar\pi_{ra})\; [23]$. Here
$\langle 0 |\psi_0^1|0\rangle =1$, see Ref. \cite{DG1}.
This amplitude is type $\phi\phi F$.
It is useful to express the invariant measures as 
\be
d\gamma_M = \prod_{r=1}^3 d\rho_r{1\over (\rho_1-\rho_2) (\rho_2-\rho_3) 
(\rho_3-\rho_1)}\qquad{\hbox{and}}\qquad d\gamma_S = 
{d\lambda^1 \over \lambda^1}
\label{imzero}\ee
in the $d=0$ sector, where $\lambda^a = \lambda^a_0$.
Comparing (\ref{ggbarc}) with (\ref{ggc}),
we verify 
the \break $d=0$ tree  $\langle A_{1}^{A_1}(\rho_1) A_{1}^{A_2}(\rho_2) 
\bar C(\rho_3) \rangle_{\rm tree}$ is the antiholomorphic version of  
the $d=1$ coupling, $\langle A_{-1}^{A_1}(\rho_1) A_{-1}^{A_2}(\rho_2) 
C(\rho_3) \rangle_{\rm tree}$.
Similarly, the $\phi\phi F$ tree
\begin{align}
\langle A_{-1}^{A_1}(\rho_1)& A_1^{A_2}(\rho_2) e_2(\rho_3) 
\rangle_{\rm tree} = \int \langle 0| A_{-1}^{A_1}(\rho_1) A_1^{A_2}(\rho_2)
e_2(\rho_3) | 0\rangle\prod_{r=1}^3 d\rho_r /d\gamma_Md\gamma_S\cr
&= i \int \prod_{r=1}^3 dk_r \,{k_1^3\over k_2k_3^2}\, \prod_{ra}
\delta(\pi^a_r-k_r\lambda^a) \langle 0|e^{\sum_{r=1}^3 
ik_r\bar\pi_{r\db}\mu^{\db}(\rho_r)}\bar\pi_3^{\dot a} Y_{\dot a}(\rho_3)
|0\rangle\cr
&\hskip40pt \times \prod_a d\lambda^a\prod_rd\rho_r/d\gamma_Sd\gamma_M\, 
\left(-{\delta^{A_1 A_2} A_{-1(1)} A_{1(2)} e_{2(3)}\over (\rho_1-\rho_2)^2
}\right)\cr
&= - \delta^{A_1A_2} [31] {(\pi_3{}^1)^2\over \pi_2{}^1} \int 
{d\lambda^2\over \lambda^1} \prod_{r=1}^3\delta(\pi_r{}^2 - 
{\lambda^2\over \lambda^1} \pi_r{}^1)\; \prod_a\delta(\sum_{r=1}^3 {\pi_r^1
\over\lambda^1} 
\bar\pi_{ra})\; A_{-1(1)} A_{1(2)} e_{2(3)}\cr
%&= - \delta^{a_1a_2}\;\delta^4(\Sigma \pi^a\bar\pi_b) \; {(\pi_1{}^1)^3\over \pi_2{}^1(\pi_3{}^1)^2} [23] [31] \, A_{-1(1)} A_{1(2)} e_{2(3)}\cr
&= -\,\delta^{A_1A_2}\;\delta^4(\Sigma \pi_r\bar\pi_r) \;  {[23]^4\over [12]^2} 
\, A_{-1(1)} A_{1(2)} e_{2(3)},
\end{align}
is the antiholomorphic version of (\ref{gge}).
The $FFF$ amplitude 
\begin{align}
\langle e_2(\rho_1) &e_2(\rho_2) \bar C(\rho_3) \rangle_{\rm tree} 
= \int \langle 0| e_2(\rho_1) e_2(\rho_2)\bar C(\rho_3) | 0\rangle
\prod_{r=1}^3 d\rho_r /d\gamma_Md\gamma_S\cr
&= -i  \int \prod_{r=1}^3 dk_r \;{k_3^2\over k_1^2 k_2^2} 
\prod_{ra}\delta(\pi_r{}^a-k_r\lambda^a) \;\langle 0| \prod_{r=1}^3\, 
e^{ik_r\bar\pi_{r\db}\mu^{\db}(\rho_r)}\,\bar\pi_r^{\dot a} Y_{\dot a}(\rho_r)
|0\rangle\cr 
&\hskip40pt \times \prod_a d\lambda^a\prod_rd\rho_r/d\gamma_Sd\gamma_M\, e_{2(1)} e_{2(2)} \bar C_{0(3)}\cr
&= -i \int \prod_{r=1}^3 dk_r \;{k_3^2\over k_1^2 k_2^2} \prod_{ra}
\delta(\pi^a_r-k_r\lambda^a(\rho_r)) \;({-i\over \lambda^1})^3\, \pi_2{}^1 
(\pi_1{}^1)^2\,{[12]^2[31] \over(\rho_1-\rho_2)(\rho_2-\rho_3)
(\rho_3-\rho_1)} \cr 
&\hskip40pt \times \prod_{a}\delta(\Sigma_{r=1}^3 k_r\bar\pi_{r a})\; 
\prod_a d\lambda^a\prod_rd\rho_r/d\gamma_Sd\gamma_M\, e_{2(1)} e_{2(2)} 
\bar C_{0(3)}\cr
&= \delta^4(\Sigma \pi_r\bar\pi_r) \, [12]^4\; e_{2(1)} e_{2(2)} \bar C_{0(3)}
\end{align}
is the antiholomorphic version of  (\ref{eec}).

Of course, we expect these results for the $d=0$ amplitudes,
from the conjugation properties of the vertex operators.
But we present the derivations to demonstrate our computational methods,
and to verify (\ref{eee}). 
The $d=0$ three-graviton coupling vanishes identically, 
since the vertex operator $e_{-2}(\rho)$ involves 
$\lambda_a(\rho) \partial \lambda^a(\rho)$ which vanishes for 
$\lambda^a(\rho) =  \lambda^a_0$, a constant:

\begin{align}
\langle e_2&(\rho_1) e_2(\rho_2) e_{-2}(\rho_3) \rangle_{\rm tree} = 
\int \langle 0| e_2(\rho_1) e_2(\rho_2) e_{-2}(\rho_3) | 0\rangle
\prod_{r=1}^3 d\rho_r / d\gamma_Md\gamma_S\cr
&= -\int \prod_{r=1}^3 dk_r \;{k_3^5\over k_1^2 k_2^2} 
%\prod_{ra} \delta(\pi^a_r-k_r\lambda^a(\rho_r)) \;
\langle 0|\prod_{r=1}^2 
e^{ik_r\bar\pi_{rb}\mu^b(\rho_r)}\,\bar\pi_r^{\dot a} Y_{\dot a}(\rho_r)\; 
e^{ik_3\bar\pi_{r\db}\mu^{\db}(\rho_3)}|0\rangle
\;\prod_rd\rho_r/d\gamma_Sd\gamma_M
\cr
&\hskip40pt \times \langle 0| \prod_{ra}
\delta(\pi^a_r-k_r\lambda^a(\rho_r))\; 
\lambda_a(\rho_3)\partial \lambda^a(\rho_3) | 0\rangle 
\;  e_{2(1)} e_{2(2)}
e_{-2(3)}= 0,
%\prod_a d\lambda^a\prod_rd\rho_r/d\gamma_Sd\gamma_M\, e_{2(1)} e_{2(2)} 
%e_{-2(3)} = 0,
\label{eeebar}\end{align}
and 
\begin{align}
\langle \bar C&(\rho_1) e_2(\rho_2) C(\rho_3) \rangle_{\rm tree} = 
\int \langle 0| \bar C(\rho_1) e_2(\rho_2) C(\rho_3) | 0\rangle
\prod_{r=1}^3 d\rho_r / d\gamma_Md\gamma_S\cr
&= - \int \prod_{r=1}^3 dk_r \;{k_1^2 k_3\over k_2^2} 
%\prod_{ra}\delta(\pi^a_r-k_r\lambda^a(\rho_r)) \;
\langle 0|\prod_{r=1}^2 
e^{ik_r\bar\pi_{ri\db}\mu^{\db}(\rho_r)}\,\bar\pi_r^{\dot a} Y_{\dot a}
(\rho_r)\; e^{ik_3\bar\pi_{r\db}\mu^{\db}(\rho_3)}|0\rangle
\;\prod_rd\rho_r/d\gamma_Sd\gamma_M\cr
&\hskip40pt \times 
\langle 0| \prod_{ra}
\delta(\pi^a_r-k_r\lambda^a(\rho_r))\,
\lambda_a(\rho_3)\partial \lambda^a(\rho_3)
|0\rangle\; \bar C_{0(1)} e_{2(2)} C_{0(3)} =0.
\label{cecbar}\end{align}
These are $FFG$ trees.
Finally, we include the familiar degree zero three-gluon vertex 
\begin{align}
\langle  A_1^{A_1}&(\rho_1) A_1^{A_2}(\rho_2) A_{-1}^{A_3}(\rho_3) 
\rangle_{\rm tree} = \int \langle 0| A_1^{A_1}(\rho_1) A_1^{A_2}(\rho_2)
A_{-1}^{A_3}(\rho_3) | 0\rangle \prod_{r=1}^3 d\rho_r / d\gamma_Md\gamma_S\cr
&= \int \prod_{r=1}^3 dk_r {k_3^3\over k_1k_2}\, \prod_{ra}
\delta(\pi^a_r-k_r\lambda^a) \;{f^{A_1A_2A_3}\over 
(\rho_1-\rho_2)(\rho_2-\rho_3)(\rho_3-\rho_1)}\cr
&\hskip40pt \times\langle 0| e^{\sum_{r=1}^3 ik_r\bar\pi_{r\db}\mu^{\db}
(\rho_r)}
|0\rangle \prod_a d\lambda^a\prod_rd\rho_r/d\gamma_Sd\gamma_M\, 
A_{1(1)} A_{1(2)} A_{-1(3)}\cr
%&= \int \prod_{r=1}^3 dk_r {k_3^3\over k_1k_2}\, \prod_{ra}\delta(\pi^a_r-k_r\lambda^a(\rho_r)) \;{f^{a_1a_2a_3}\over (\rho_1-\rho_2)(\rho_2-\rho_3)(\rho_3-\rho_1)}\cr
%&\hskip40pt \times \prod_a d\lambda^a d\mu^a \; e^{i\sum_{r=1}3 k_r\bar\pi_{rb}\mu^b(\rho_r)}\quad \prod_rd\rho_r/d\gamma_Sd\gamma_M\quad A_{1(1)} A_{1(2)} A_{-1(3)}\cr 
%%
%&= \delta^4(\Simga\pi_r\bar\pi_r)\, [23] \,f^{a_1a_2a_3}\; {(\pi_3{}^1)^3\over (\pi_1{}^1)^2 \pi_2{}^1}\, A_{1(1)} A_{1(2)} A_{-1(3)}\cr
%%
&= \delta^4(\Sigma \pi_r\bar\pi_r)\, f^{A_1A_2A_3}\; {[12]^3\over [23][31]} 
\,A_{1(1)} A_{1(2)} A_{-1(3)}.
\end{align}
\begin{center}
\renewcommand{\tabcolsep}{1cm}
\renewcommand{\arraystretch}{1.4}
\vskip20pt\begin{tabular}{| l |} 
\hline
$\langle A^{A_1}_{1}A^{A_2}_{1} \bar C \rangle = 
-[12]^2\delta^{A_1A_2} A_{1(1)}A_{1(2)}\bar C_{0(3)}
\delta^4(\Sigma \pi_r \bpi_r)$ \\ \hline
$\langle A_{-1}^{A_1}A_{1}^{A_2}e_{2} \rangle = -
{[23]^4\over  [12]^2} \delta^{A_1A_2} A_{-1(1)}A_{1(2)} e_{2(3)}
\delta^4(\Sigma \pi_r \bpi_r)$
\\ \hline
$\langle e_{2}e_{2} \bar C \rangle = [12]^4
e_{2(1)}e_{2(2)}\bar C_{0(3)}\delta^4(\Sigma \pi_r \bpi_r)$
\\ \hline
$\langle e_{2}e_{2}e_{-2} \rangle = 0$\\ \hline
$\langle \bar C e_{2} C \rangle = 0$\\ \hline
$\langle A^{A_1}_{1}A^{A_2}_{1}A^{A_3}_{-1} \rangle = {[12]^3 \over
[23][31]} f^{A_1 A_2 A_3} A_{1(1)}A_{1(2)}A_{-1(3)}
\delta^4(\Sigma \pi_r \bpi_r)$\\ \hline
\end{tabular}
\centerline{\hskip-20pt Table 5: 
$d=0$ Unprimed conformal supergravity couplings}
\end{center}
\vskip 10pt

\subsection{\sl Amplitudes with Primed Vertices}

In this section we will compute tree amplitudes containing states with
primed vertex operators. As a preliminary study, consider the $d=1$ 
coupling 
$\langle A_{-1}^{A_1}(\rho_1)A_{-1}^{A_2}(\rho_2)C'(\rho_3) \rangle_{\rm tree}$
with $\phi\phi G'$ vertex operators.
Using the previous methods, it is convenient to evaluate the primed coupling as 
\begin{align}
&\langle A_{-1}^{A_1}(\rho_1)A_{-1}^{A_2}(\rho_2)C'(\rho_3) \rangle_{\rm tree}
= \int \langle 0| e^{q_0}A_{-1}^{A_1}(\rho_1)A_{-1}^{A_2}(\rho_2)C'(\rho_3)
|0\rangle \prod_{r=1}^3 d\rho_r / d\gamma_Md\gamma_S\cr
&= -\int \prod_{r=1}^3 dk_r \prod_{r=1}^3 k_r \,
\prod_{ra}
\delta({\pi_r}^a-k_r\lambda^a(\rho_r)) \,
k_1^2 k_2^2 \,(\rho_1-\rho_2)^2 \,\prod_{r=1}^3 d\rho_r \,
\prod_a d^2\lambda^a / d\gamma_Md\gamma_S \delta^{A_1 A_2} \cr
&\hskip5truemm\times 
\langle 0|e^{q_0} \,i \left( {s_{3\dot a}\over k_3}
\partial \mu^{\dot a}(\rho_3) - s_{3\dot a}
\mu^{\dot a}(\rho_3) \bar s_{3a}
\partial\lambda^a(\rho_3) \right)\;
e^{i\sum_{r=1}^3 k_r\bar\pi_{r\db}\mu^{\db}(\rho_r)}|0\rangle
\;\; A_{-1(1)} A_{-1(2)} C'_{0(3)}   \cr
&= -\prod_{r=1}^3 {\pi_r^1\over(\lambda^1(\rho_r))^2}\;
\delta\left(\pi^2_r - {\lambda^2(\rho_r)\over\lambda^1(\rho_r)} \pi_r^1\right)
(\rho_1-\rho_2)^2 
({\pi_1^1\pi_2^1\over\lambda^1(\rho_1)\lambda^1(\rho_2)})^2\cr
&\hskip25pt \times
\; \prod_{a,\da} d^2\lambda^a d^2\mu^{\da} \prod_rd
\rho_r/d\gamma_Sd\gamma_M\,\delta^{A_1 A_2} \; A_{-1(1)} A_{-1(2)} C'_{0(3)}\cr
&\hskip35pt \times i \Big( {s_{31}\over\pi_3^1}\;
\left(\lambda^1_0\mu^1_{-1} 
- \mu^1_0\lambda^1_{-1}\right)
+ {s_{32}\over\pi_3^2}\;\left( \lambda^2_0\mu^{2}_{-1} 
- \mu^2_0\lambda^2_{-1}\right)\Big)
e^{i\sum_{r=1}^3 {\pi_r^1\over\lambda^1(\rho_r)}\bar\pi_{r\db}
(\mu^{\db}_0 + \rho_r\mu^{\db}_{-1})}\cr
\label{eeprime}
\end{align}
where we have used the delta functions
$\delta(\pi_r^1 - k_r \lambda^1(\rho_r))$ to do the $k_r$ integrations.
Here $\lambda^a(\rho) = \lambda_0^a + \rho\lambda_{-1}^a$.
In order to perform the $d^2\mu^{\da}$ integrations, we note that
\be\sum_{r=1}^n\pi_r^b\bar\pi_{r\da}=\sum_{r=1}^n{\lambda^b(\rho_r){\pi_r}^1
\bar\pi_{r\da}\over
\lambda^1(\rho_r)}=\lambda^b_0\sum_{r=1}^n{\pi_r^1\bar\pi_{r\da}\over
\lambda^1(\rho_r)}
+\lambda_{-1}^b\sum_{r=1}^n{\pi_r^1\bar\pi_{r\da}\rho_r \over\lambda^1(\rho_r)}
\label{pbp}\ee
for any $n$, when $\pi_r^2-(\lambda^2(z_r)/\lambda^1(z_r))\pi_r^1=0$. 
We can invert this change of variables to write 
$\sum_{r=1}^2 {\pi_r^1\over\lambda^1(\rho_r)}\bar\pi_{r\da}$
and
$\sum_{r=1}^2 {\pi_r^1\over\lambda^1(\rho_r)}\bar\pi_{r\da} \rho_r$
in terms of $\sum_{r=1}^2 \pi^b\bar\pi_{\da}$, and express the
exponential in  (\ref{eeprime}) as
\be  
e^{i\sum_{r=1}^3 {\pi_r^1\over\lambda^1(\rho_r)}\bar\pi_{r\db}
(\mu^{\db}_0 + \rho_r\mu^{\db}_{-1})} =
e^{i{\epsilon_{ca}\over\det\lambda} 
\left(\lambda_0^a\mu_{-1}^{\db} - \lambda_{-1}^a\mu_0^{\db}\right)
\sum_{r=1}^3 \pi_r^c\bar\pi_{r\db}},
\ee
where the anti-symmetric epsilon tensor is
$\epsilon^{12} = 1 = -\epsilon_{12}$, as in section 2.
Then the integrand of the $d^2\mu^{\da}$ integrations can be expressed as 
derivatives of the exponential,  

\begin{align}
&\int \prod_{\da} d^2\mu^{\da}
\Big( {s_{31}\over\pi_3^1}\;
\left(\lambda^1_0\mu^1_{-1}
- \mu^1_0\lambda^1_{-1}\right)
+ {s_{32}\over\pi_3^2}\;\left( \lambda^2_0\mu^{2}_{-1}
- \mu^2_0\lambda^2_{-1}\right)\Big)
e^{i\sum_{r=1}^3 {\pi_r^1\over\lambda^1(\rho_r)}\bar\pi_{r\db}
(\mu^{\db}_0 + \rho_r\mu^{\db}_{-1})}\cr
&= (-i \det\lambda)\,
\left(  {s_{31}\over\pi_3^1}\; {\partial\over\partial\sum_{r=1}^3
\pi_r^2\bpi_{r1}} - 
{s_{32}\over\pi_3^2}\;{\partial\over\partial\sum_{r=1}^3
\pi_r^1\bpi_{r2}}\right)
\int  \prod_{\da} d^2\mu^{\da}
e^{i{\epsilon_{ca}\over\det\lambda}
\left(\lambda_0^a\mu_{-1}^{\db} - \lambda_{-1}^a\mu_0^{\db}\right)
\sum_{r=1}^3 \pi_r^c\bar\pi_{r\db}}.
\label{der}
\end{align}
Performing the $d^2\mu^{\da}$ integrals
to find momentum delta functions,
\begin{align}
\int  \prod_{\da} d^2\mu^{\da}
e^{i{\epsilon_{ca}\over\det\lambda}
\left(\lambda_0^a\mu_{-1}^{\db} - \lambda_{-1}^a\mu_0^{\db}\right)
\sum_{r=1}^3 \pi_r^c\bar\pi_{r\db}}
= (\det\lambda)^2 \,\delta^4(\Sigma\pi_r\bpi_r),
\end{align} 
and using our previous
methods, (\ref{eeprime}) becomes 
\begin{align}&
\langle
A_{-1}^{A_1}(\rho_1) A_{-1}^{A_2}(\rho_2)
C'(\rho_3) \rangle_{\rm tree}\cr
&=  - \langle 12\rangle^2
\Big[ {s_{31}\over\pi_3^1}\,
\delta^2(\sum_{r=1}^3 \pi_r^a\bar\pi_{r2})\,
\delta(\sum_{r=1}^3
\pi^1_r\bar\pi_{r1})\,
\delta'\left(\sum_{r=1}^3 \pi^2_r\bar\pi_{r1} \right)\cr
&\hskip45pt - {s_{32}\over\pi_3^2}\,
\delta^2(\sum_{r=1}^3 \pi_r^a\bar\pi_{r1})\,
\; \delta(\sum_{r=1}^3 \pi^2_r\bar\pi_{r2})
\;\delta'\left(\sum_{r=1}^3
\pi^1_r\bar\pi_{r2}\right)
\Big]\;
\delta^{A_1 A_2} A_{-1(1)} A_{-1(2)} C'_{0(3)}\cr
&= \langle 12\rangle^2 \,
\delta^{A_1 A_2} A_{-1(1)} A_{-1(2)} {C'_{0(3)}
\over 2p_3^0} {\partial\over \partial P^0} 
\delta^4(\Sigma \pi_r\bar\pi_r)\,
\label{eeprimea}\end{align}
where we  have chosen the Berkovits Witten gauge $s_{\dot a} =
{\pi^a\sigma^0_{a\dot a}\over 2 p^0}$, so
${s_{21}\over\pi_2^1}= {s_{22}\over\pi_2^2} = {1\over \pi_2^1\bar\pi_2^1
+\pi_2^2\bar\pi_2^2} = {1\over 2 p_2^0}$, and defined
$P^0 = \sum_{r=1}^3 p^0_r = 
\half \sum_{r=1}^3 (\pi^1_r\bar\pi_{r2} - \pi_r^2\bpi_{r1})$,
using $p_{ra\da} = \pi_{ra}\bpi_{r\da} = \sigma^\mu_{a\da}p_{r\mu}$
as in section 2. 

We interpret the amplitude (\ref{eeprimea}) 
with the help of understanding how the momentum operator acts on
the primed states. 
In conformal supergravity, the dipole pairs arise as solutions to 
equations of motion with higher than quadratic derivatives, 
see for example \cite{FZ, BW}. Each pair $\sigma_p,\sigma'_p$ 
satisfies $(\partial_\mu\partial^\mu)^2
\sigma = 0$, and comprises a plane wave state
$\sigma_p = e^{ip\cdot x}$, and a state 
$\sigma'_p = i A\cdot x  e^{ip\cdot x}$ that cannot diagonalize
the momentum operator 
for any non-zero vector $A$ independent of $x$. 
Since $P^{\rm{op}}_{a\da} = -i{\partial\over\partial x^{a\da}}$,
then 
\be P^{\rm{op}}_{a\da} \sigma_p = p_{a\dot a} \sigma_p, \qquad
P^{\rm{op}}_{a\da} \sigma'_p = p_{a\dot a} \sigma'_p
+ A_{a\da} \sigma_p. 
\label{mo}\ee
In particular, we can write
$\sigma'_p 
=
A^{a\da}{\partial\over\partial p^{a\da}} \sigma_p,$
and choose $A$ to be in the time direction \cite{BW}
to make contact with the Berkovits Witten gauge, so
%and write
%the general solution for the scalar dipole as
%\be C(x) = \int d^4p \delta(p^2) e^{ip\cdot x} \left( C_0(p)
%+ {ix_0\over 2p_0} C_0'(p)\right).
%\ee
\be
\sigma'_p \sim {\partial\over\partial p^0} \sigma_p.
\label{wf}\ee

The primed amplitude (\ref{eeprimea})
is effectively $-{C'_{0(3)}\over 2 p_3^0C_{0(3)}}{\partial\over\partial p_3^0}$
times the form (\ref{ggc}), as expected in view of (\ref{wf}).
For the pair of states to have the same relative dimension,
the wavefunctions $C_{0(r)}, C'_{0(r)}$
differ in dimension by a factor of $(p^0)^2$, 
so the primed amplitude (\ref{eeprimea}) has canonical dimensions.
%To compare more easily, we define $\tilde C_{0(r)}$ to have the same
%dimension as $C_{0(r)}$,
%\be C'_{0(r)} = 2(p^0_r)^2 \tilde C_{0(r)}.
%\ee

But what about momentum conservation? Surely primed amplitudes 
are conformally invariant, just as the others. 
Although the primed states are not eigenstates of the momentum operator,
we know they transform as in (\ref{mo}). 
So, the momentum operator acts on the coupling
$\langle
A_{-1}^{A_1}(\rho_1) A_{-1}^{A_2}(\rho_2)
C'(\rho_3) \rangle_{\rm tree}$ as
\begin{align}
& P^0 
\langle
A_{-1}^{A_1}(\rho_1) A_{-1}^{A_2}(\rho_2)
C'(\rho_3) \rangle_{\rm tree}
- {C'_{0(3)}\over C_{0(3)} 2p_3^0}\,
\langle
A_{-1}^{A_1}(\rho_1) A_{-1}^{A_2}(\rho_2)
C(\rho_3) \rangle_{\rm tree}\cr
& =
\langle 12\rangle^2 \,\delta^{A_1 A_2} A_{-1(1)} A_{-1(2)} {C'_{0(3)}\over
2p_3^0}
\;\, P^0{\partial\over \partial P^0} \delta(P^0) \delta^3(P^i)\cr
&\hskip100pt + {C'_{0(3)}\over 2p_3^0}\,
\langle 12\rangle^2 \, \delta^{A_1 A_2} A_{-1(1)} A_{-1(2)}
\delta(P^0) \delta^3(P^i)\cr &= 0
\end{align}
and
\begin{align}
& P^i\langle
A_{-1}^{A_1}(\rho_1) A_{-1}^{A_2}(\rho_2)
C'(\rho_3) \rangle_{\rm tree}
=
\langle 12\rangle^2 \,
\delta^{A_1 A_2} A_{-1(1)} A_{-1(2)}
\; C'_{0(3)}{\partial\over \partial P^0} \delta(P^0) P^i \delta^3(P^i)
=0,\end{align}
verifying the primed amplitude
(\ref{eeprimea}) has translational invariance. Here $P^\mu = 
\sum_r p^\mu_r$, and $P^0{\partial\over\partial P^0} \delta(P^0) = 
-\delta(P^0)$ on the support of a test function.
\vskip10pt 

In a similar calculation, now using the $e_{-2}'(\rho)$ vertex operator
in lieu of $C'(\rho)$, we find the MHV coupling for two gluons and
a primed graviton:
\begin{align}
&\langle
A_1^{A_1}(\rho_1) A_{-1}^{A_2}(\rho_2)
e_{-2}'(\rho_3) \rangle_{\rm tree} 
= \int \langle 0|
e^{q_0} A_1^{A_1}(\rho_1) A_{-1}^{A_2}(\rho_2)e_{-2}'(\rho_3) | 0\rangle
\prod_{r=1}^3 d\rho_r / d\gamma_Md\gamma_S\cr
&=- {\langle 23\rangle^4\over \langle 12\rangle^2}
\delta^{A_1 A_2} A_{1(1)} A_{-1(2)} e'_{-2(3)}
\Big[ {s_{31}\over\pi_3^1}\,{\partial\over\partial\sum_{r=1}^3
\pi_r^2\bpi_{r1}} -
{s_{32}\over\pi_3^2}\;{\partial\over\partial\sum_{r=1}^3
\pi_r^1\bpi_{r2}}\Big]\;
\delta^4(\sum_{r=1}^2 \pi_r\bar\pi_r)\cr
&= {\langle 23\rangle^4\over\langle 12\rangle^2} \,
\delta^{A_1 A_2} A_{1(1)} A_{-1(2)} {e'_{-2(3)}
\over 2p_3^0} {\partial\over \partial P^0}
\delta^4(\Sigma \pi_r\bar\pi_r).
\end{align}
\vskip10pt 
For the $d=1$ $GGG$ coupling of two gravitons and a scalar, 
we can extend to any combination of primed vertices $V_{G'}$ as follows. 
If there is more than one primed vertex operator,
there will be a product of factors in the derivation of the 
amplitude, of the form,
\be \Big( {s_{r1}\over\pi_r^1}\;
\left(\lambda^1_0\mu^1_{-1}
- \mu^1_0\lambda^1_{-1}\right)
+ {s_{r2}\over\pi_r^2}\;\left( \lambda^2_0\mu^{2}_{-1}
- \mu^2_0\lambda^2_{-1}\right)\Big)
\ee
for each site $r$ that corresponds to a primed vertex. 
We can evaluate this in a similar way to (\ref{der}),
to find, for example, 
\begin{align}
\langle
e'_{-2}(\rho_1) e'_{-2}(\rho_2)
C'(\rho_3) \rangle_{\rm tree} &= \langle 0|
e^{q_0} e'_{-2}(\rho_1) e'_{-2}(\rho_2) C'(\rho_3) | 0\rangle
\prod_{r=1}^3 d\rho_r / d\gamma_Md\gamma_S
\cr
&= -\langle 12\rangle^4 \,
{e'_{-2(1)}\over 2p_1^0}{e'_{-2(2)}\over 2p_2^0} {C'_{0(3)}
\over 2p_3^0} {\partial^3\over (\partial P^0)^3}
\delta^4(\Sigma \pi_r\bar\pi_r).
\label{sequence}
\end{align}
So effectively, the contribution of a $V_{G'}(\rho_r)$ vertex operator
to a tree amplitude can be found by replacing {\it each} unprimed wavefunction
by a primed wavefunction times $-{1\over 2p_r^0}{\partial\over\partial
p_r^0}$.

Amplitudes involving $V_F'$ vertices are more tedious to evaluate.
As a guide for these methods, we can use the antiholomorphic amplitudes.
For example, the $d=0$ three-point coupling

\begin{align}
&\langle A_1^{A_1}(\rho_1) A_1^{A_2}(\rho_2) \bar C'(\rho_3) \rangle_{\rm tree} 
=\int \langle 0| A_1^{A_1}(\rho_1) A_1^{A_2}(\rho_2)\bar C'(\rho_3) 
| 0\rangle \prod_{r=1}^3 d\rho_r/d
\gamma_Sd\gamma_M\cr
&=-\int \prod_{r=1}^3 dk_r {k_3^2\over k_1k_2}\;\prod_{r=1}^3 d\rho_r/d
\gamma_Sd\gamma_M\; \left( {\delta^{A_1 A_2} A_{1(1)} A_{1(2)} \bar C'_{0(3)}
\over (\rho_1-\rho_2)^2}\right) \cr
&\hskip10pt\times\big(\;\bar s_3^a \;\; \langle 0|\prod_{c, r=1}^2 
\delta(\pi_r^c - k_r\lambda^c(\rho_r)) \;\delta(\pi_3^c - k_3\lambda^c(\rho_r))
\; Y_a(\rho_3) |0\rangle \, \int \prod_a d\mu^a_0 
e^{i\sum_{r=1}^3 k_r\bar\pi_{rb}\mu^b_0}\cr
&\hskip30pt + is_3^{\dot a} \,\,\langle 0|e^{i\sum_{r=1}^2 k_r \bar
\pi_{r\dot b} \mu^{\dot b}(\rho_r)} e^{i k_3 \bar\pi_{3\dot b}
\mu^{\dot b}(\rho_3)}\, Y_{\dot a}(\rho_3) |0\rangle \cr
&\hskip50pt \times
\int\prod d\lambda_0^a\prod_{c,r=1}^2 \delta(\pi_r^c - k_r\lambda^c(\rho_r))\; 
\bar s_3^a {\partial\over \partial \pi_3^a}  
\delta(\pi_3^c - k_3\lambda^c(\rho_3))\big)\cr
&=[12]^2 \,
\delta^{A_1 A_2} A_{1(1)} A_{1(2)} {\bar C'_{0(3)}
\over 2p_3^0} {\partial\over \partial P^0}
\delta^4(\Sigma\pi_r\bar\pi_r).
\end{align}
where we could evaluate
$\langle 0|\prod_{c, r=1}^2 \delta(\pi_r^c - k_r\lambda^c(\rho_r))
\,\delta(\pi_3^c - k_3\lambda^c(\rho_r))\; Y_a(\rho_3) |0\rangle
$ by writing the delta functions $\prod_{c, r} \delta(\pi_r^c - k_r
\lambda^c(\rho_r))$ as
$\int\prod_{c, r} d\omega_{rc} e^{i\sum_{r=1}^2 \omega_{rc} (k_r
\lambda^c(\rho_r) - \pi_r^c)}$, using the commutator of $\lambda^a(\rho_r)$
with $Y_a(\rho_3)$, for $r=1,2$, and divide by the invariant measure
(\ref{imzero}). But we know the result, since it is the antiholomorphic
form of (\ref{eeprimea}), found by replacing 
$\pi_{ra},\bpi_{r\db}$ 
with their conjugates $\bpi_{r\da},\pi_{rb}$.

The $d=0$ three-point amplitudes from the $FGG'$, $F'GG$, $GGG'$, $FG'G'$,
$F'GG'$, $ GG'G'$, $F'G'G'$ and $G'G'G'$ sectors all vanish. 
For $d=0$, the three-point tree amplitudes from the $FFG'$ and $FGF'$ sectors
give conventional Einstein couplings, along the lines of \cite{BrWu}.
Three-point trees in the $FF'G'$, $F'F'G$ and $F'F'G'$ sectors for $d=0$
are more detailed to access.

The MHV three-point amplitudes involving the primed states
of the dipoles in the $\phi\phi G'$,
$GGG'$, $GG'G'$ and $G'G'G'$ sectors, together
with their helicity conjugates are summarized in Tables 6 and 7.

\begin{center}
\renewcommand{\tabcolsep}{1cm}
\renewcommand{\arraystretch}{2.2}
\vskip20pt
\begin{tabular}{| l |} \hline
%{\bf Primed MHV Couplings} \\ \hline
$\langle A^{A_1}_{-1}A^{A_2}_{-1}C'\rangle = 
\langle 12 \rangle^2\,\delta^{A_1A_2} A_{-1(1)}A_{-1(2)}{C'_{0(3)}\over 2p_3^0}
{\partial\over \partial p_3^0}\delta^4(\Sigma \pi_r\bpi_r)$
\\ \hline
$\langle A_1^{A_1}A_{-1}^{A_2}e'_{-2} \rangle = 
{\langle 23 \rangle^4\over \langle 12 \rangle^2} \delta^{A_1A_2}
A_{1(1)}A_{-1(2)}{e'_{-2(3)}\over 2p_3^0}{\partial\over \partial p_3^0}
\delta^4(\Sigma \pi_r \bpi_r)$\\ \hline
$\langle e_{-2}e_{-2}C'\rangle = -
\langle 12 \rangle^4  
e_{-2(1)}e_{-2(2)}{C'_{0(3)}\over 2p_3^0}{\partial\over \partial p_3^0}
\delta^4(\Sigma \pi_r \bpi_r)$\\ \hline
$\langle e_{-2}e'_{-2}C\rangle = -
\langle 12 \rangle^4
e_{-2(1)}{e'_{-2(2)}\over 2p_2^0}C_{0(3)}{\partial\over \partial p_2^0}
\delta^4(\Sigma \pi_r \bpi_r)$\\ \hline
$\langle e_{-2}e'_{-2}C'\rangle = -
\langle 12 \rangle^4  
e_{-2(1)}{e'_{-2(2)}\over 2p_2^0} 
{C'_{0(3)}\over 2p_3^0}
{\partial^2\over\partial p_2^0\partial p_3^0}
\delta^4(\Sigma \pi_r \bpi_r)$\\ \hline
$\langle e'_{-2}e'_{-2}C\rangle = -
\langle 12 \rangle^4
{e'_{-2(1)}\over 2p_1^0}
{e'_{-2(2)}\over 2p_2^0} 
C_{0(3)}{\partial^2\over \partial p_1^0\partial p_2^0}
\delta^4(\Sigma \pi_r \bpi_r)$\\ \hline
$\langle e'_{-2}e'_{-2}C'\rangle = -
\langle 12 \rangle^4
{e'_{-2(1)}\over 2p_1^0}
{e'_{-2(2)}\over 2p_2^0} {C'_{0(3)}\over 2p_3^0}
{\partial^3\over \partial p_1^0\partial p_2^0
\partial p_3^0}
\delta^4(\Sigma \pi_r \bpi_r)$\\ \hline
\end{tabular}
\vskip10pt
\centerline{Table 6: MHV Conformal supergravity couplings 
with primed states}
\end{center}
\vskip 30pt

\begin{center}
\renewcommand{\tabcolsep}{1cm}
\renewcommand{\arraystretch}{2.2}
\vskip20pt
\begin{tabular}{| l |} \hline
%{\bf Primed dzero Couplings} \\ \hline
$\langle A^{A_1}_1A^{A_2}_1\bar C'\rangle =
[12]^2 \,\delta^{A_1A_2} A_{1(1)}A_{1(2)}{\bar C'_{0(3)}\over 2p_3^0}
{\partial\over \partial p_3^0}\delta^4(\Sigma \pi_r\bpi_r)$
\\ \hline
$\langle A_{-1}^{A_1}A_{1}^{A_2}e'_{2} \rangle =
{[23]^4\over [12]^2} \delta^{A_1A_2}
A_{-1(1)}A_{1(2)}{e'_{2(3)}\over 2p_3^0}{\partial\over \partial p_3^0}
\delta^4(\Sigma \pi_r \bpi_r)$\\ \hline
$\langle e_{2}e_{2}\bar C'\rangle = -
[12]^4
e_{2(1)}e_{2(2)}{\bar C'_{0(3)}\over 2p_3^0}{\partial\over \partial p_3^0}
\delta^4(\Sigma \pi_r \bpi_r)$\\ \hline
$\langle e_{2}e'_{2}\bar C\rangle = -
[12]^4
e_{2(1)}{e'_{2(2)}\over 2p_2^0}\bar C_{0(3)}{\partial\over \partial p_2^0}
\delta^4(\Sigma \pi_r \bpi_r)$\\ \hline
$\langle e_{2}e'_{2}\bar C'\rangle = -
[12]^4
e_{2(1)}{e'_{2(2)}\over 2p_2^0}
{\bar C'_{0(3)}\over 2p_3^0}
{\partial^2\over \partial p_2^0\partial p_3^0}
\delta^4(\Sigma \pi_r \bpi_r)$\\ \hline
$\langle e'_{2}e'_{2}\bar C\rangle = -
[12]^4
{e'_{2(1)}\over 2p_1^0}
{e'_{2(2)}\over 2p_2^0}
\bar C_{0(3)}{\partial^2\over \partial p_1^0\partial p_2^0}
\delta^4(\Sigma \pi_r \bpi_r)$\\ \hline
$\langle e'_{2}e'_{2}\bar C'\rangle = -
[12]^4
{e'_{2(1)}\over 2p_1^0}
{e'_{2(2)}\over 2p_2^0} {\bar C'_{0(3)}\over 2p_3^0}
{\partial^3\over \partial p_1^0\partial p_2^0
\partial p_3^0}
\delta^4(\Sigma \pi_r \bpi_r)$\\ \hline
\end{tabular}
\vskip10pt
\centerline{Table 7: $d=0$ Conformal supergravity couplings 
with primed states}
\end{center}

\vfill\eject
%%%
% SECTION
%%%
\section{Canonical Derivation of the Berkovits Witten Amplitudes}

In this section, we extend our analysis of the three-point functions
using canonical quantization, to $N$-point MHV tree amplitudes 
for unprimed vertex operators.
The maximal helicity violating (MHV) amplitudes contain any two vertex operators
of negative helicity, $e_{-2}, \bar{C}, A_{-1}$, 
and $N-2$ positive helicity vertex operators from the set $e_{2}, C, A_1$.

\par We will compute an amplitude for a specific choice of the two
negative helicity states, and then discuss how this generalizes.
We consider the $(d=1)$ amplitude for two negative helicity vertex operators, 
one of type $G$ and one of type $F$, 
$\langle e_{-2}\bar{C}e_{2}\ldots e_{2}C 
\ldots C A_1 \ldots A_1 \rangle_{\hbox{tree}}$. This has 
$n$ positive helicity type $F$ gravitons, 
$m$ positive helicity type $G$ scalars, and $p$ positive helicity gluons .  
We denote the total number of vertices as $N=2+n+m+p$.  
Inserting the expressions from Table 3, we find
\begin{align}
\int \langle 0| e^{q_0} &\int dk_1 k_1 \lambda_{a_1}(\rho_1) \prod_{a=1}^{2} 
\delta (k_1\lambda^a(\rho_1) - \pi_1{}^a)e^{ik_1\bar{\pi}_{1 \dot{b}}
\mu^{\dot{b}}(\rho_1)} k_1^4 \psi^1(\rho_1)\psi^2(\rho_1)\psi^3(\rho_1)
\psi^4(\rho_1) \partial \lambda^{a_1}(\rho_1)  \cr
&\times i \int \frac{dk_2}{k_2^2} \bar{\pi}_2{}^{\dot{a}} \prod_{a=1}^{2} 
\delta (k_2\lambda^a(\rho_2) - \pi_2{}^a) e^{ik_2\bar{\pi}_{2\dot{b}}
\mu^{\dot{b}}(\rho_2)} k_2^4 \psi^1(\rho_2)\psi^2(\rho_2)\psi^3(\rho_2)
\psi^4(\rho_2) Y_{\dot{a}} (\rho_2) \cr
& \times \prod^{n+2}_{j=3} i \int \frac{dk_j}{k_j^2} \bar{\pi}_j{}^{\dot{a}} 
\prod_{a=1}^{2} \delta (k_j\lambda^a(\rho_j) - \pi_j{}^a) 
e^{ik_j\bar{\pi}_{j\dot{b}}\mu^{\dot{b}}(\rho_j)} Y_{\dot{a}} (\rho_j) \cr
& \times \prod^{m+n+2}_{j=n+3} \int dk_j k_j \lambda_{a_j}(\rho_j) 
\prod_{a=1}^{2} \delta (k_j\lambda^a(\rho_j) - \pi^a_j) 
e^{ik_j\bar{\pi}_{j\db}\mu^{\dot{b}}(\rho_j)} 
\partial \lambda^{a_j}(\rho_j) \cr
&\times \prod^{N}_{j=m+n+3} \int \frac{dk_j}{k_j}\prod_{a=1}^{2}
\delta(k_j \lambda^a(\rho_j) - \pi^a_j) e^{ik_j\bar{\pi}_{j\db}
\mu^{\dot{b}}(\rho_j)} J^{A_j}(\rho_j) |0\rangle 
\prod_{r=1}^N d\rho_r / d\gamma_S d\gamma_M
\label{nmp}\end{align}
where we have dropped the polarizations for convenience.
It is useful to introduce the sets of indices:  
${\bf n} = \{3,  \ldots , n+2\}$, ${\bf m} = \{n+3, \ldots , m+n+2\}$, 
and ${\bf p} = \{m+n+3, \ldots , N\}$.  To further emphasize the occurrence 
of gluon or graviton type, we define the larger sets ${\bf n}' 
= \{2, 3,  \ldots , n+2\}$ and ${\bf m}' = \{1, n+3, \ldots , m+n+2\}$.
%\footnote{This example requires no need to define an increased set ${\bf p}'$,
%however in some other MHV calculations it will be useful to form this set as 
%well.}.  
From the following formula presented below, 
we can see these sets will be useful 
when considering amplitudes having a more complicated ordering of 
vertex operators.  
We rewrite (\ref{nmp}) as

\begin{align}
\left(i\right)^{n+1} & \int \prod_{r=1}^N dk_r d\rho_r / d\gamma_S d\gamma_M\,
\prod_{a,r} 
\delta(k_r \lambda^a(\rho_r) - \pi_r{}^a) \prod_{j \in {\bf n}'}
\left(\frac{1}{k_j}\right)^2 \prod_{j \in {\bf m}'} 
\left(k_j\right) \prod_{j \in {\bf p}} \left(\frac{1}{k_j}\right) \cr
& \times \left( k_1k_2\right)^4 \langle 0|e^{q_0} \psi^1(\rho_1)
\psi^1(\rho_2)\psi^2(\rho_1)\psi^2(\rho_2)\psi^3(\rho_1)\psi^3(\rho_2)
\psi^4(\rho_1) \psi^4(\rho_2)|0\rangle \cr
& \times \langle 0| e^{q_0} \prod_{j \in {\bf m}'}
\lambda_{a}(\rho_j) \partial \lambda^{a}(\rho_j) |0\rangle 
\; \langle 0| \prod_{j \in{\bf p}} J^{A_j} (\rho_j) |0\rangle \cr
&\times\langle 0| e^{q_0} e^{ik_1\bar{\pi}_{1\db}\mu^{\dot{b}}(\rho_1)} 
e^{ik_2\bar{\pi}_{2 \db}\mu^{\db}(\rho_2)}
\bar{\pi}_2{}^{\dot{a}} 
Y_{\dot{a}}(\rho_2)  \prod_{j \in{\bf n}} \left(e^{ik_j\bar{\pi}_{j\db}
\mu^{\db}(\rho_j)}\bpi_j^{\da}Y_{\da}(\rho_j)\right)  \cr
& \hspace{2cm} \times \prod_{j \in {\bf m}} \left( e^{ik_j\bar{\pi}_{j \db}
\mu^{\db}(\rho_j)}\right)
\prod_{j \in {\bf p}} \left( e^{ik_j\bar{\pi}_{j \db}\mu^{\db}(\rho_j)}\right)
|0\rangle.\label{nmptwo}
\end{align}

Many simplifications happen at this stage.  With 
\begin{equation}
\langle 0 | e^{q_0}\psi^1(\rho_1)\psi^1(\rho_2)|0\rangle = 
(\rho_1 - \rho_2) \langle 0| e^{q_0} \psi^1_{-1}\psi^1_0 |0\rangle
= (\rho_1- \rho_2),
\end{equation}
four factors of $\rho_1-\rho_2$ come from the second line.  
Evaluating the $\lambda$ term,  we find 
\begin{equation}
\langle 0| e^{q_0} \prod_{j \in {\bf m}'} \lambda_{a_j}(\rho_j) 
\partial\lambda^{a_j}(\rho_j) |0\rangle = \int\prod_a d^2\lambda^a
(\det \lambda)^{m+1},
\end{equation}
where $\det \lambda = \lambda^1_0 \lambda^2_{-1} -\lambda^2_0 \lambda^1_{-1}$,
as in section 3.. 
We use a current algebra contribution \cite{FZh}
\begin{equation}
\langle 0| \prod_{j \in{\bf p}} J^{A_j} (\rho_j) |0\rangle = f^{A_{m+n+3} 
\cdots A_N}\prod_{j \in {\bf p}}\frac{1}{\rho_j - \rho_{j+1}},
\label{fzh}\end{equation}
with $\rho_{N+1} \equiv \rho_{m+n+3}$.  In what follows, we
denote $f^{A_{m+n+3} \ldots A_N}=f^{A \ldots A}$.  
This is merely simplification of notation, as the group indices add no 
new information not contained in the denominator.  We note, for computing
MHV amplitudes containing negative helicity gluons,
the form (\ref{fzh})
remains the same with the set $\bf p$ replaced by the
total set of gluons $\bf p'$.

\par The last expectation value in (\ref{nmptwo}) is equal to 
\begin{equation}
\left(-i\right)^{n+1} (\det \lambda)^2 \delta^4\left(\Sigma \pi_r 
\bar{\pi}_r\right) \prod_{x \in {\bf n}'}\sum^N_{y=1,y \neq x} 
k_y \frac{[xy]}{\rho_y - \rho_x},
\label{withy}\end{equation}
where now $\delta^4\left(\Sigma \pi_r\bar{\pi}_r\right) \equiv 
\prod_{\da,b} \delta\left(\Sigma_{r=1}^N \pi_r^b\bpi_{r\da}\right)$.
We integrate the $k_r$'s using $\delta(k_r - \pi^1_r / \lambda^1(\rho_r))$, 
and evaluate the amplitude (\ref{nmptwo}) to obtain 
\begin{align}
\delta^4&(\Sigma\pi_r\bpi_r)  \int \prod_{r=1}^N d\rho_r \prod_a d^2 
\lambda^a / d\gamma_S\gamma_M \prod_{r=1}^N \delta(\pi_r{}^2 - 
\frac{\lambda^2(\rho_r)}{\lambda^1(\rho_r)}\pi_r{}^1) \prod_{r=1}^N \frac{1}
{\lambda^1(\rho_r)} \cr
& \times \left(\frac{\pi_1{}^1}{\lambda^1(\rho_1)} \frac{\pi_2{}^1}
{\lambda^1(\rho_2)} \right)^4(\rho_1 - \rho_2)^4 \prod_{j \in {\bf n}'}  
\left( \frac{\lambda^1(\rho_j)}{\pi^1{}_j}\right)^2 \prod_{j \in {\bf m}'} 
\left( \frac{\pi_j{}^1{}}{\lambda^1(\rho_j)}\right) \prod_{j \in {\bf p}} 
\left( \frac{\lambda^1(\rho_j)}{\pi_j{}^1}\right)   \cr
& \times f^{A \cdots A} \prod_{j \in {\bf p}} \left( \frac{1}{\rho_j - 
\rho_{j+1}} \right) \left(\det \lambda\right)^{m+3}\prod_{x \in {\bf n}'}
\sum^N_{y=1,y \neq x} \frac{\pi_y{}^1}{\lambda^1(\rho_y)} \frac{[xy]}
{\rho_y - \rho_x}\label{nmpfour}\end{align}
We define $\zeta_r = \frac{\lambda^2(\rho_r)}{\lambda^1(\rho_r)}$ and 
change variables from $\rho_r$ to $\zeta_r$. The identification 
\begin{equation}
\sum_{y=1,y \neq x}^N \frac{\pi_y{}^1}{\lambda^1(\rho_y)}\frac{[xy]}
{\rho_y - \rho_x} = \frac{\det \lambda}{(\lambda^1(\rho_x))^3}
\sum^N_{y=1, y\neq x}\frac{\pi_y{}^1[xy]}{\zeta_y - \zeta_x}
\label{chv}\end{equation}
follows from
\be\sum_{y\ne x}{\pi^1_y [xy]\over\lambda^1(\rho_y)(\rho_y-\rho_x)}
={1\over\lambda^1(\rho_x)}\sum_{y\ne x} {\pi^1_y [xy]\over(\rho_y-\rho_x)},
\ee
using $\sum_y{\pi_y^1\bpi_{y\db}\over\lambda^1(\rho_y)} = 0$, which is
provided by the factor $\delta^4(\Sigma\pi_r\bpi_r)$ in (\ref{nmpfour}),
in view of the equality (\ref{pbp}).
To implement the change of variables, we have
$\zeta_r-\zeta_j={(\rho_r-\rho_j)\det\lambda \over\lambda^1(\rho_r)
\lambda^1(\rho_j)},\;$
$d\zeta = {\det\lambda \over\lambda^1(\rho)^2}d\rho$, 
so (\ref{nmpfour}) is
\begin{align}
\delta^4&(\Sigma \pi_r\bar{\pi}_r)\int \prod_{r=1}^N d\zeta_r \prod_a 
d^2 \lambda^a / d\gamma_S d\gamma_M \left(\det \lambda\right)^{-2} 
\prod_{r=1}^N 
\delta(\pi_r{}^2 - \zeta_r \pi_r^1) \left(\pi_1{}^1\pi_2{}^1
(\zeta_1 - \zeta_2)\right)^4\cr
&\hskip-3pt\times\prod_{j \in {\bf n}'}  \left( \frac{1}{\pi_j{}^1}\right)^2 
\prod_{j \in {\bf m}'} \left(\pi_j{}^1\right) \prod_{j \in {\bf p}}  
\left( \frac{1}{\pi_j{}^1}\right) f^{A \cdots A}\prod_{j \in {\bf p}} 
\left( \frac{1}{\zeta_j - \zeta_{j+1}} \right) \prod_{x \in {\bf n}'} 
\sum_{y = 1, y\neq x}^N \pi_y{}^1\frac{[xy]}{\zeta_y - \zeta_x}
\end{align}
We identify $d\gamma_S d\gamma_M = d^2\lambda^a (\det \lambda)^{-2}$ and 
do the $\zeta_r$ integrations. Since 
\begin{equation}
\prod_{x \in {\bf n}'} \left( \frac{1}{\pi_x{}^1} \right)^3 
\sum_{y=1, y\neq x}^N\pi_y{}^1 \frac{[xy]}{\zeta_y - \zeta_x} = 
\prod_{x \in {\bf n}'}\sum_{y=1, y\neq x}^N\frac{(\pi_y^1)^2 }
{(\pi_x^1)^2} \frac{[xy]}{\langle xy \rangle},
\label{nfa}\end{equation}
(\ref{nfa}) can be reexpressed as \cite{BW}
\begin{equation}
\prod_{x \in {\bf n}'}\sum_{y=1, y\neq x}^N\frac{\langle y \xi\rangle ^2}
{\langle x \xi \rangle^2} \frac{[xy]}{\langle xy \rangle},
\end{equation}
which is independent of $\pi_\xi$,$\bpi_\xi$,
to obtain the result
\begin{align}
\langle e_{-2}\bar{C} e_{2} \cdots e_{2} C & \cdots C A_1 \ldots A_1
\rangle \cr 
& = \delta^4 (\Sigma \pi_r\bar{\pi}_r)\langle 12 \rangle^4 f^{A \cdots A}
\prod_{j \in {\bf p}}\frac{1}{\langle j, j+1 \rangle} \;\prod_{i\in {\bf n}'} 
\sum^N_{j=1, j\ne i}
\frac{\langle j \xi \rangle^2}{\langle i \xi \rangle^2}\frac{[ij]}
{\langle ij \rangle}.
\label{can}\end{align}
The amplitude is independent of the order of the positive helicity 
states, as this corresponds merely to changing the position of the
$Y_{\da}$ fields, and does not affect (\ref{withy}). 
To generalize our expression for any two negative helicity states, 
it is useful to identify pieces common to all amplitudes: 
the two negative helicity states in any position $\rho_r,\rho_s$
will give the factor of $\langle rs \rangle^4$, 
all gluon vertices contribute to $ f^{A \cdots A}\prod_{j \in {\bf p}'}
\frac{1}{\langle j, j+1 \rangle}$, defined in (\ref{fzh}),
and the type $F$  vertex operators 
contribute to the 
product of sums. The type $G$ 
vertex operators provide factors of $\det\lambda$, 
and leave no further mark on the amplitude.  
Our answer (\ref{can}) thus becomes the Berkovits Witten formula \cite{BW},
which they found from a path integral formulation, and 
where we have absorbed a factor $(-i)^F$ in the definition of
the vertex operators $V_F$.
\vskip30pt

\leftline{\sl Comparison of Conformal Gravity with Einstein Gravity Amplitudes}

To visualize conformal gravity amplitudes better, we use (\ref{can})
to study the conformal four-graviton tree amplitude 

\begin{align}
&\langle e_{-2}(\rho_1) e_{-2}(\rho_2) e_2(\rho_3) e_2(\rho_4)
\rangle_{CG} =
\langle 12\rangle^4 \prod_{j=3,4}\sum_{k\ne j} {[jk] \langle k\xi\rangle^2
\over \langle jk\rangle \langle j\xi\rangle^2}\cr
&=-{\langle 12\rangle^4
[32]\langle 21\rangle \left ( \langle 43\rangle \langle 21\rangle
-\langle 23\rangle \langle 41\rangle\right) \, [42]\langle 21\rangle
\left( \langle 34\rangle \langle 21\rangle
- \langle 24\rangle \langle 31\rangle\right)\over
\langle 31\rangle^2 \langle 41\rangle^2
\langle 34\rangle^2 \langle 23\rangle \langle 42\rangle}
\qquad {\hbox{(choose $\xi =1$)}}\cr
&= \, {\langle 12\rangle^4 [34]^4\over (s_{12})^2}
\qquad {\hbox{using the identity 
$\langle 43\rangle\langle 21\rangle -
\langle 23\rangle\langle 41\rangle = \langle 13\rangle \langle 24\rangle$}}
\cr
&= {s_{23} s_{24}\over s_{12}}\;
\langle e_{-2}(1) e_{-2}(2) e_2(3) e_2(4)\rangle_{Einstein},
\label{fgrav}\end{align}
{\it which has fewer poles} than Einstein gravity,
since the Berends Giele Kuijf expression \cite{Berends} for Einstein gravity 
tree amplitudes as a product of Yang-Mills trees is

\begin{align}
\langle e_{-2}(1) e_{-2}(2) e_2(3) e_2(4)\rangle_{Einstein}
&= s_{12} {\langle 12\rangle^3\over\langle 23\rangle
\langle 34\rangle\langle 41\rangle}
\; {\langle 12\rangle^3\over\langle 24\rangle
\langle 43\rangle\langle 31\rangle}\cr
%= {\langle 12\rangle^8 [12]\over
%\langle 12\rangle \langle 23\rangle\langle 34\rangle\langle 41\rangle
%\langle 24\rangle\langle 13\rangle\langle 34\rangle}\cr
%&=-{s_{12}^2\over s_{23} s_{24}}
%\;\left( {\langle 12\rangle [34]\over [12]\langle 34\rangle}\right)^2
%= s_{12} \cdot {\langle 12\rangle^2 [34]^2\over s_{12} s_{23}}
%{\langle 12\rangle^2 [34]^2\over s_{12} s_{24}} =
&= {1\over s_{12} s_{23} s_{24}} \langle 12\rangle^4[34]^4.\nonumber
%\quad {\hbox{ see for eg. hep-th/9802162, eq (2.23).}}\nonumber
\end{align}

We can reproduce the conformal gravity four-point function (\ref{fgrav}) 
from tree level exchange of the scalar field C
\begin{align}
\langle e_{-2}(\rho_1) e_{-2}(\rho_2) e_2(\rho_3) e_2(\rho_4)
\rangle_{CG} &=
\langle 12\rangle^4 {1\over (s_{12})^2} [34]^4
\end{align}
corresponding to the product of the three-point trees $\;$
$\langle e_{-2}(\rho_1) e_{-2}(\rho_2) C(\rho)\rangle = \langle 12\rangle^4$
and \break $\langle \bar C(\rho) e_2(\rho_3) e_2(\rho_4) \rangle = [34]^4$,
times the conformal propagator ${1\over (p^2)^2}$, where $p^2 = s_{12}.$

\vskip30pt
%%%
% SECTION
%%%
\section{$N$-point Tree Amplitudes for Mixed Primed and Unprimed Vertices}

\par Finally we turn to the $N$-point MHV scattering amplitudes containing 
both primed and unprimed vertices. To begin, consider two 
negative helicity gluons, $A_{-1}$ and $n$ $G'$ scalars, $C'_{0}$.  
The total number of vertices is $N=2+n$. 
The set of primed vertices is ${\bf n} = \{3, \dots , N \}$.  
To compute
$\langle A^{A_1}_{-1}(\rho_1)A^{A_2}_{-1}(\rho_2)C'_{0}(\rho_3) 
\ldots C'_{0}(\rho_N) 
\rangle_{\rm{tree}}$, use the vertex operators from Table 3, 
\hskip-10pt
\begin{align}
\int \langle 0 |e^{q_0}&\int \frac{dk_1}{k_1} \prod_a\delta
\left( k_1 \lambda^a(\rho_1) - \pi_1{}^a \right) e^{i k_1 \bar{\pi}_{1 \db} 
\mu^{\db}(\rho_1)} k_1^4 \psi^1(\rho_1) \psi^2(\rho_1) 
\psi^3(\rho_1) \psi^4 (\rho_1) A_{-1(1)} J^{A_1}(\rho_1)
\cr
& \times \int \frac{dk_2}{k_2} \prod_a \delta \left( k_2 
\lambda^a(\rho_2) - \pi_2{}^a \right) e^{i k_2 \bar{\pi}_{2 \db} 
\mu^{\db}(\rho_2)} k_2^4 \psi^1(\rho_2) \psi^2(\rho_2) \psi^3(\rho_2) 
\psi^4 (\rho_2) A_{-1(2)} J^{A_2}(\rho_2) \cr 
&\times \prod_{j \in \bn} \big ( i \int dk_j k_j\prod_a \delta 
\left( k_j \lambda^a(\rho_j) - \pi^a{}_j\right) e^{i k_j \bar{\pi}_{j\db} 
\mu^{\db}(\rho_j)}\cr
&\hskip50pt \times \left[s_{j\dot{a}}\partial\mu^{\dot{a}}(\rho_j) - 
\bar{s}_{ja} s_{j \dot{a}}\mu^{\dot{a}}(\rho_j)\partial \lambda^a(\rho_j)
\right] C'_{0(j)}\big)| 0\rangle  \prod_{r=1}^N d\rho_r/d
\gamma_Sd\gamma_M,
\end{align}
which yields
\begin{align}
i^n \int & \prod_{r=1}^N dk_r  d\rho_r k_r \prod_{a,r} \delta 
\left( k_r \lambda^a(\rho_r) -\pi_r{}^a \right) 
\left( \rho_1-\rho_2\right)^4(k_1k_2)^4\prod_a d^2\lambda^a 
/ d\gamma_S d\gamma_M \cr
&\times  \frac{-\delta^{A_1A_2}}{(\rho_1 - \rho_2)^2}  
\left( \frac{1}{k_1k_2} \right)^2 A_{-1(1)}A_{-1(2)}C'_{0(3)} \cdots C'_{0(N)} 
\cr
&\times \langle 0 | e^{q_0} \prod_{j \in \bn} 
\left( \frac{s_{j\dot{a}}}{k_j}\partial \mu^{\dot{a}}
(\rho_j) - \bar{s}_{ja}s_{j\dot{a}}\mu^{\dot{a}}(\rho_j) 
\partial \lambda^a(\rho_j) \right)e^{i\sum_r k_r \bar{\pi}_{r\db}
\mu^{\dot{b}}(\rho_j)} | 0 \rangle.
\label{allone}\end{align}
We evalute the expectation value in (\ref{allone}) as a sequence of
derivatives, as in (\ref{sequence}),
and find
\begin{align}
&\langle A^{A_1}_{-1}(\rho_1)A^{A_2}_{-1}(\rho_2)C'_{0}(\rho_3)
\ldots C'_{0}(\rho_N)
\rangle_{\rm{tree}}\cr
&= -\langle 12 \rangle^2 \delta^{A_1A_2}
A_{-1(1)}A_{-1(2)}C'_{0(3)} \ldots C'_{0(N)} \prod_{j \in \bn} 
\left[ -{1\over 2p^0_j}{\partial\over\partial p^0_j} 
\right] \delta^4(\Sigma\pi_r \bar{\pi}_r ).
\end{align}
Clearly this same form holds for
\begin{align}
&\langle e'_{-2}(\rho_1)e'_{-2}(\rho_2)C'_{0}(\rho_3)\ldots C'_{0}(\rho_N)
\rangle_{\rm{tree}}\cr
&= \langle 12 \rangle^4 
e'_{-2(1)}e'_{-2(2)}C'_{0(3)} \ldots C'_{0(N)} \prod_{j \in N}
\left[ -{1\over 2p^0_j}{\partial\over\partial p^0_j}
\right] \delta^4(\Sigma\pi_r \bar{\pi}_r ),
\end{align}
and for any combination of these type $G$ primed and unprimed states,
with the product then taken over the primed sites.

$N$-point functions with type $F'$ vertices, and with mixed
$G'$ and $F$  are more varied to track.
%For the four-point function, we can use the anti-holmorphic form
%of 
%$\langle e'_{-2}(\rho_1)e'_{-2}(\rho_2)e_2(\rho_3)e_2(\rho_4)
%\rangle_{\rm{tree}}$
%to evaluate the $F'F'GG$ amplitude 
%$\langle e'_2(\rho_1)e'_2(\rho_2)e_{-2}(\rho_3)e_{-2}(\rho_4)
%\rangle_{\rm{tree}}$.

%END SECTION
\vskip50pt

\section*{Aknowledgements}
We are grateful to Peter Goddard for important contributions in the 
early stages
of this work, and Michael Chesterman, Lance Dixon, and Edward Witten
for valuable comments. LD thanks 
the Institute for Advanced Study at Princeton for its hospitality.
LD and JI were partially supported by the U.S. Department of Energy,
Grant No. DE-FG01-06ER06-01, Task A.
\vfill\eject

\singlespacing

%%% References %%%

\providecommand{\bysame}{\leavevmode\hbox to3em{\hrulefill}\thinspace}
\providecommand{\MR}{\relax\ifhmode\unskip\space\fi MR }
% \MRhref is called by the amsart/book/proc definition of \MR.
\providecommand{\MRhref}[2]{%
  \href{http://www.ams.org/mathscinet-getitem?mr=#1}{#2}
}
\providecommand{\href}[2]{#2}

\end{document}